\documentclass[sigconf]{acmart}

\usepackage{amsmath,amsfonts}
\usepackage{algorithm}
\usepackage{graphicx}
\usepackage{array}            % ← 必须：提供 \newcolumntype
\usepackage{subfig}           % 建议: \usepackage[caption=false,font=footnotesize]{subfig}
\usepackage{algpseudocode}
\usepackage{makecell}
\usepackage{multirow}
\usepackage{tikz}
\AtBeginDocument{%
  }

\definecolor{myyellow}{RGB}{255,217,101}
\definecolor{myblue}{RGB}{102,153,255}
\definecolor{mygreen}{RGB}{146,208,80}

\newcommand*\circled[1]{%
  \tikz[baseline=(char.base)]{
    \node[text=white, circle, draw, fill=black, inner sep=0.3pt] (char) {#1};}}

\setcopyright{acmlicensed}
\copyrightyear{2026}
\acmYear{2026}
\acmDOI{XXXXXXX.XXXXXXX}
\acmConference[ICS'26]{July 06--09,
  2026}{Belfast, Northern Ireland, UK}
\begin{document}

\title{Accelerating Multi-Scale Deformable Attention Using Near-Memory-Processing Architecture}

\author{Huize Li}
%\email{webmaster@marysville-ohio.com}
\affiliation{%
  \institution{University of Central Florida}
  \city{Orlando}
  \state{Florida}
  \country{USA}
}
\email{huize.li@ucf.edu}

\author{Qinggang Wang}
\affiliation{%
  \institution{Huazhong University of Sci. \& Tech.}
  \city{Wuhan}
  \country{China}}
\email{qgwang@hust.edu.cn}

\author{Bin Gao}
\affiliation{%
  \institution{National University of Singapore}
  \city{Singapore}
  \country{Singapore}
}
\email{bingao@comp.nus.edu.sg}

\author{Dan Chen}
\affiliation{%
  \institution{National University of Singapore}
  \city{Singapore}
  \country{Singapore}
}
\email{danchen@nus.edu.sg}

\author{Yu Huang}
\affiliation{%
  \institution{Huazhong University of Sci. \& Tech.}
  \city{Wuhan}
  \country{China}}
\email{yuh@hust.edu.cn}

\author{Xin Xin}
\authornote{Corresponding author is Xin Xin.}
%\email{webmaster@marysville-ohio.com}
\affiliation{%
  \institution{University of Central Florida}
  \city{Orlando}
  \state{Florida}
  \country{USA}
}
\email{xin.xin@ucf.edu}

\begin{abstract}
Multi-Scale Deformable Attention (MSDAttn) has become a fundamental component in various vision tasks due to its effective multi-scale grid sampling (MSGS). However, its reliance on random sampling results in highly irregular memory access patterns, making it a memory-intensive operation inefficient for GPUs. Near-memory processing (NMP) offers a promising solution for accelerating memory-bound kernels, yet existing NMP-based attention accelerators remain suboptimal for MSDAttn due to incompatible load balancing and data reuse strategies. Specifically, current NMP solutions uniformly distribute processing elements (PEs) across all banks, leading to significant PE underutilization and excessive cross-bank data transfers. Moreover, most rely on locality-based reuse, which fails under MSDAttn’s unpredictable sampling patterns.

To address these challenges, this paper presents \textit{DANMP}, a hardware–software co-designed NMP-based MSDAttn accelerator. On the hardware side, \textit{DANMP} adopts non-uniform NMP integration to handle unbalanced workloads, allocating PEs only in select banks for hot entries, while cold data are processed at the bank-group level—reducing PE idleness and cross-bank transfers. On the software side, it introduces a clustering-and-packing (CAP) method that leverages clustering to improve temporal locality in query processing, enhancing data reuse. Finally, we implement host–NMP co-optimization techniques, including an optimized programming model, customized instructions, and a tailored dataflow. Experiments on object detection inference show that \textit{DANMP} achieves 97.43$\times$ speedup and 208.47$\times$ energy efficiency improvement over NVIDIA A6000 GPU.

\end{abstract}




\maketitle

\section{Introduction}

The \textit{Detection Transformer} (DETR)~\cite{carion2020end, gupta2022ow, beal2020toward, Fang2021} has recently achieved remarkable success in object detection by combining \textit{convolutional neural networks} (CNNs) with a Transformer-based encoder-decoder architecture~\cite{Vaswani17}. Despite its improvement in accuracy, DETR suffers from slow convergence, requiring significantly more training epochs than conventional object detection methods~\cite{du2018understanding, xie2021oriented, chen2016r}. {To address this limitation, \textit{multi-scale deformable attention} (MSDAttn/MSDeformAttn) has been introduced, effectively reducing training/inference costs and computational complexity.} Inspired by deformable convolution~\cite{Dai2017ICCV}, MSDAttn selectively attends to a sparse set of reference points across multi-scale feature maps, avoiding the exhaustive computation inherent in traditional attention mechanisms~\cite{Vaswani17}. By leveraging efficient \textit{multi-scale grid-sampling} (MSGS), MSDAttn achieves state-of-the-art training/inference efficiency. However, the irregular distribution of reference point sampling introduces random memory access overhead, significantly constraining system performance~\cite{xu2024defa}.

Recently, there has been growing interest in hardware-software co-designed solutions to mitigate random memory access overhead. These solutions encompass various platforms, including GPUs~\cite{zhu2021deformable, longformer20, Tay20, Rajpurkar18}, \textit{Field Programmable Gate Arrays} (FPGAs)~\cite{Zhang21fpga, Fan2022fpga, bai2024dac, Zhang21}, and \textit{Application Specific Integrated Circuits} (ASICs)~\cite{xu2024defa, Qu22, Qin23isca, Lu21}. While these co-designed approaches outperform software-only solutions with hardware-friendly scheduling and computation techniques, they remain compute-centric and are still constrained by off-chip memory bandwidth when accelerating memory-intensive MSDAttn workloads. As a result, they suffer from high-latency off-chip accesses due to the large volume of randomly sampled feature maps~\cite{xu2024defa}. Given the increasing demand for low latency object detection, improving the efficiency of MSDAttn could have a substantial impact on both performance and scalability.

\begin{figure}[t]
\centering
\includegraphics[width=8.3cm]{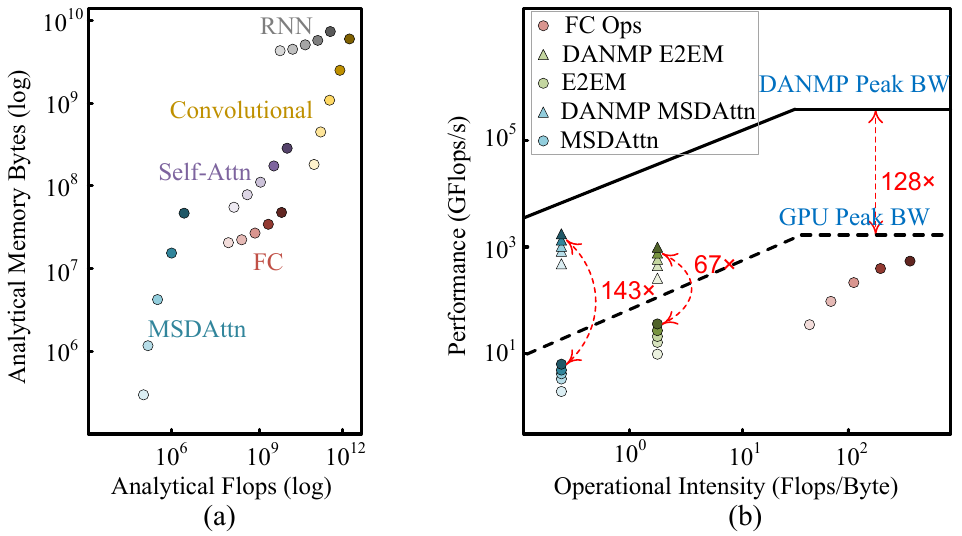}
\vspace{-1em}
\caption{(a) Computational and memory footprint of common deep learning operators across different batch sizes (darker shades indicate larger batches), (b) {Bandwidth roofline} model showing performance gains, including operator-level speedups for FC and MSDAttn, as well as end-to-end model (E2EM) acceleration enabled by \textit{DANMP}}
\label{cm_ratio}
%\vspace{-2em}
\end{figure}

\begin{figure}[t]
\centering
\includegraphics[width=8.3cm]{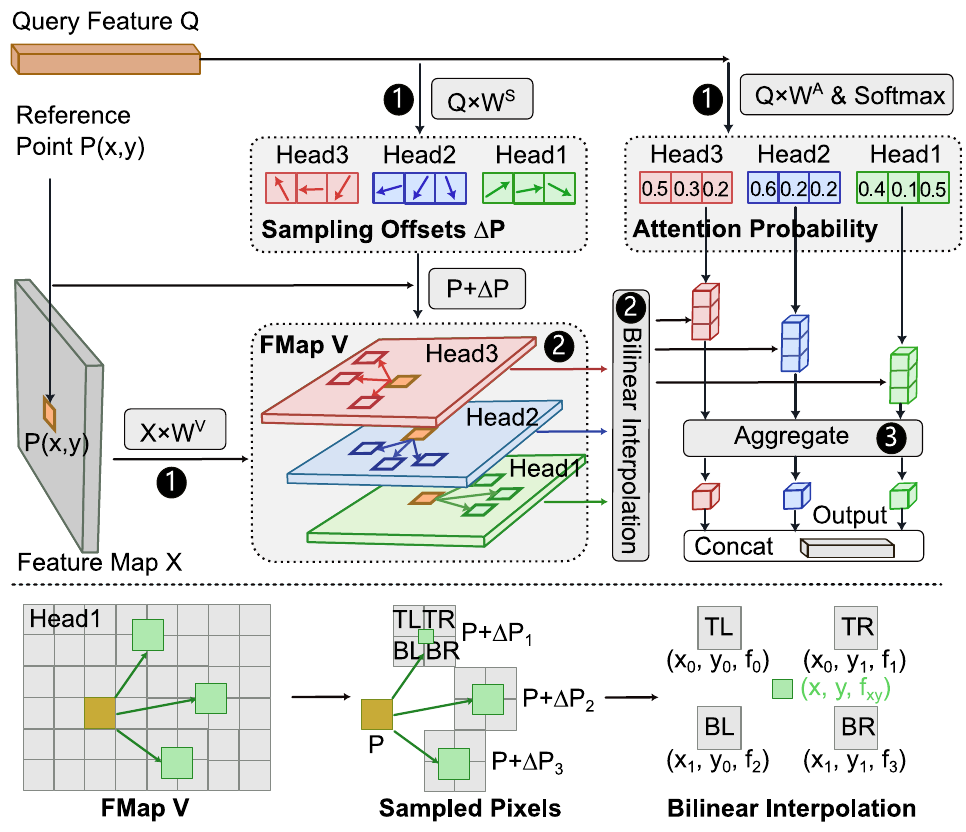}
\vspace{-1em}
\caption{Overview of the MSDAttn}
\label{msattention}
\vspace{-1em}
\end{figure}

A quantitative comparison of the computational and memory access requirements for various neural network operations is presented in Figure~\ref{cm_ratio}(a) ({The experimental configuration is same to Section~\ref{motivation_ana}}). Unlike \textit{recurrent neural networks} (RNNs), convolutions, \textit{full-connection} (FC), and \textit{self-attention} (Self-Attn) operations, {MSDAttn exhibits exceptionally low operational intensity---less than 10\% of the compute–bandwidth intersection in the Roofline model---indicating that its performance is primarily constrained by memory bandwidth rather than arithmetic throughput. This memory-bound behavior arises mainly from multi-scale grid sampling (\circled{2} in Figure~\ref{msattention}) and feature aggregation (\circled{3} in Figure~\ref{msattention}).}

Compared to standard self-attention, MSDAttn faces three major challenges.
Irregular Sampling: Sparse and non-uniform sampling across large feature maps leads to unpredictable memory accesses, making prefetching and dataflow optimizations ineffective.
Limited Data Reuse: Hundreds of detection queries target different spatial regions, offering little locality across queries.
Irregular Sparsity and Load Imbalance: MSDAttn prunes data-dependent feature points, creating highly unstructured sparsity and severe load imbalance.
{Unlike previous sparsity accelerators (e.g., SCNN~\cite{parashar2017scnn}, CANDLES~\cite{gudaparthi2022candles}, AWB-GCN~\cite{geng2020awb}, EnGN~\cite{liang2020engn}) that exploit predictable zero patterns, MSDAttn’s query-specific sparsity varies across scales and tokens, breaking regular dataflows and tile reuse. Consequently, its operational intensity remains extremely low, leaving the computation memory-bound.}

This paper presents \textit{DANMP}, the first near-memory processing architecture designed to accelerate MSDAttn in DETR. \textit{DANMP} is a hardware-software co-designed \textit{Dual In-line Memory Module} (DIMM)-based system, built on standard DRAM technology. In contrast to stack-based solutions, which demands specialized 2.5D/3D integration processes such as HBM~\cite{Ahn2015isca, Zhou2022, Ding2023, micro2024li, highp2025} and ReRAM~\cite{ReGNN2022, asadi2024, resqm2020, cpsaa2024, resma2022, splim2025}, the adoption of a DIMM-based NMP~\cite{chen2023, sadimm2025, Kim2022, micro2024ham} enables the use of commodity DDR5 devices and provides large memory capacity (over 64GB) at low cost. By eliminating the off-chip memory bottleneck and exposing higher internal bandwidth, \textit{DANMP} significantly enhances both performance and energy efficiency. Specifically, it raises the {bandwidth roofline} of GPU platform by 128$\times$ in bandwidth-constrained regions (Figure~\ref{cm_ratio}(b)), unlocking optimization opportunities that are infeasible with conventional architectures.

We conducted a detailed characterization of the open-source Deformable DETR (DE-DETR) model~\cite{zhu2021deformable} as a case study (Figure~\ref{cm_ratio} and Figure~\ref{case1}). This analysis quantifies the potential benefits of near-memory processing in accelerating Deformable DETR, specifically highlighting how \textit{DANMP} can optimize performance. In this paper, memory-intensive MSDAttn operations are executed within memory, while compute-intensive FC operations are processed by the host. Additionally, we also performed an NMP-based DETR case study (Figure~\ref{case2}) to identify performance bottlenecks in executing DETR on existing NMP accelerators. Our findings reveal two key challenges that severely limit performance. \textit{PE Idleness Caused by Load Imbalance:} Current NMP solutions often adopt uniform PE integration, leading to idle PEs when mapping irregular MSDAttn workloads. Reassigning tasks to idle PEs introduces significant cross-bank transfers, further increasing memory access overhead. \textit{Bandwidth Wastage Due to Random Memory Access:} Due to the poor data locality in MSDAttn, existing NMP accelerators fail to exploit data reuse efficiently. This leads to underutilization of the ample intra-bank bandwidth and intensifies the pressure on the limited cross-bank bandwidth.

To address these challenges, \textit{DANMP} introduces both hardware and software innovations. On the hardware side, \textit{DANMP} investigates the hierarchical structure of DRAM memory systems and exploits the multi-tiered parallelism across ranks, bank-groups, and banks. Unlike existing NMP architectures, we adopt a non-traditional, uneven integration strategy by implementing PEs only in selected banks. Mapping irregular MSDAttn workloads to the unevenly integrated architecture significantly reduces PE idleness without introducing cross-bank transfers. On the software side, we introduce a \textit{clustering-and-packing} (CAP) method, which significantly enhances data reuse by grouping queries with the same sub-targets. This optimization mitigates random memory access inefficiencies and improves temporal locality. Compared to the GPU platform, \textit{DANMP} achieves 143$\times$ reduction in MSDAttn latency and 67$\times$ improvement in end-to-end DE-DETR inference performance, as shown in Figure~\ref{cm_ratio}(b). This work makes the following key research contributions:

\begin{itemize}
    \item We demonstrate that MSDAttn is constrained by memory bandwidth with our GPU case study. Our NMP-based case study reveals that load imbalance and low data reuse rates are the primary factors behind this constraint.
    \item We introduce \textit{DANMP}, a lightweight DDR5-compatible near-memory processing accelerator designed for MSDAttn execution, achieving 97.43$\times$ performance improvement, 208.47$\times$ energy efficiency improvement against NVIDIA A6000 GPU.
    \item We also explore uneven PE integration, clustering-and-packing algorithm, and host-NMP co-optimizations to enhance the performance of \textit{DANMP}.
\end{itemize}

\section{Background}
\subsection{Multi-Scale Deformable Attention}
\label{msda}
{
Multi-Scale Deformable Attention (MSDAttn) is a key operation in modern object detectors such as Deformable DETR, where each query dynamically samples a small set of spatial locations across multi-scale feature maps to capture long-range dependencies. 
This data-dependent sampling provides high modeling flexibility but also introduces severe irregularity in memory access, as illustrated in Figure~\ref{msattention}.} MSDAttn identifies relations between each query and a small set of sampling points in multi-scale fmaps $X \in \mathbb{R}^{n\times d}$. Let $n$ and $d$ denote the length of flattened feature maps from $l$ levels ($N = \sum_{i=1}^{l}H_i\times W_i$) and hidden dimension of pixel vectors respectively. Where $H_i$ and $W_i$ denote the height (number of vertical pixels) and width (number of horizontal pixels) of layer $l$. The computation of MSDAttn is calculated by concatenate all $h$ attention heads as equation~(\ref{concat}),

\vspace{-1em}
\begin{equation}
{MSDAttn}(\boldsymbol{Q}, \boldsymbol{P}, \boldsymbol{X}) = Concat(\boldsymbol{H}_1, ... , \boldsymbol{H}_h),
\label{concat}
\end{equation}
the calculation of the output matrix in the $j^{th}$ attention head is shown in equation~(\ref{head_j}),
\begin{equation}
\boldsymbol{H}_j = Softmax(\boldsymbol{Q\cdot W}_j^A) \boldsymbol{\cdot V}_j(\boldsymbol{X},\, {\boldsymbol{P} \oplus \Delta\boldsymbol{P}_j}),
\label{head_j}
\end{equation}
where $\boldsymbol{Q} \in \mathbb{R}^{n\times d}$ and $\boldsymbol{V} = \boldsymbol{X\cdot W}^V \in \mathbb{R}^{n\times d}$ denote the query matrix and the multi-scale fmaps, respectively. $\boldsymbol{P}$ refers to the coordinates of the reference point in each level of fmaps. 
{
$\oplus$ denotes coordinate indexing rather than arithmetic addition, 
i.e., $\boldsymbol{P} \oplus \Delta\boldsymbol{P}_j$ represents sampling locations offset from reference points $\boldsymbol{P}$ by learned displacements $\Delta\boldsymbol{P}_j$. 
$\boldsymbol{V}_j(\boldsymbol{X}, \boldsymbol{P} \oplus \Delta\boldsymbol{P}_j)$ gathers feature values from $\boldsymbol{X}$ via bilinear interpolation at these locations.
$\Delta\boldsymbol{P} = \boldsymbol{Q\cdot W}^S$ denotes the offsets of sampling points. $\boldsymbol{W}^A \in \mathbb{R}^{d\times h\times l\times p}$,}
while $\boldsymbol{W}^V \in \mathbb{R}^{d\times d}$ and $\boldsymbol{W}^S \in \mathbb{R}^{d\times 2n\cdot l\cdot p}$ are learnable weights, where $p$ denotes the fixed number of sampling points in each level of fmaps. {
Although the above equations are expressed in matrix form for compactness, they do not represent dense matrix multiplications. 
Instead, MSDAttn follows a gather–interpolate–accumulate pattern visualized in Figure~\ref{msattention}, which exposes its highly irregular memory behavior.}

%\begin{figure}[t]
%\centering
%\includegraphics[width=8.3cm]{figures/graph/dram.pdf}
%\vspace{-0.5em}
%\caption{Hierarchy architecture of DRAM}
%\label{dram}
%\vspace{-2em}
%\end{figure}

Figure~\ref{msattention} illustrates the visualization process of single-level deformable attention, which is an example of all levels in MSDAttn. The inputs to Deformable DETR include the query feature $Q$, reference point $P(x,y)$, feature map $X$, and all learnable weight matrices. The linear transformations on query $Q$ and feature map $X$ are performed in \circled{1}, involves three fully connected layers (weighted matrix multiplication). Simultaneously, a softmax operation normalizes all points across different levels, which are projected from a row vector of the input query, generating attention probability vectors. In \circled{2}, the \textit{multi-scale grid sampling} (MSGS) procedure applies a \textit{bilinear interpolation} (BI) kernel to process fractional sampling points, retrieving corresponding values from the multi-scale feature maps. In \circled{3}, each pixel vector obtained from sampling is multiplied by an element of the attention probability vector, and all weighted vectors are summed to produce the output of a single attention head. Finally, the outputs from all attention heads are concatenated to generate the final output matrix.

{The bottom of Figure~\ref{msattention} illustrates the MSGS computation. Three sampling points, $P+\Delta P_1$, $P+\Delta P_2$, and $P+\Delta P_3$, are selected on the feature map $V$. Since these points typically lie between discrete pixels, bilinear interpolation is used to estimate their feature values. For a sampling point at $(x, y)$ with interpolated value $f_{xy}$, its four neighboring pixels are: top-left (TL) at $(x_0, y_0)$ with value $f_0$, top-right (TR) at $(x_0, y_1)$ with $f_1$, bottom-left (BL) at $(x_1, y_0)$ with $f_2$, and bottom-right (BR) at $(x_1, y_1)$ with $f_3$. The interpolated value is computed as: $f_{xy} \approx [f_0(x_1-x)(y_1-y)+f_1(x_1-x)(y-y_1)+f_2(x-x_1)(y_1-y)+f_3(x-x_1)(y-y_1)]/[(x_1-x_0)(y_1-y_0)]$. This process involves substantial random memory access due to the irregular distribution of sampling points $P$ and $\Delta P$.}

%{{\bf Comparison with Deformable Convolution.} While both MSDAttn~\cite{zhu2021deformable} and deformable convolution~\cite{Dai2017ICCV} use dynamic, data-dependent sampling, they differ significantly in structure and flexibility. Deformable convolution applies offsets relative to a fixed kernel grid (e.g., $3\times3$) and typically shares them across channels, resulting in constrained, localized sampling with predictable access patterns. In contrast, MSDAttn enables each query to attend to multiple arbitrary positions across the feature map, with sampling determined dynamically per query and per head. It also incorporates multi-scale features and non-local learned sampling, leading to sparse and irregular memory access. These characteristics make MSDAttn far more challenging to accelerate using conventional deformable convolution hardware.}

\begin{figure}[t]
\centering
\includegraphics[width=8.39cm]{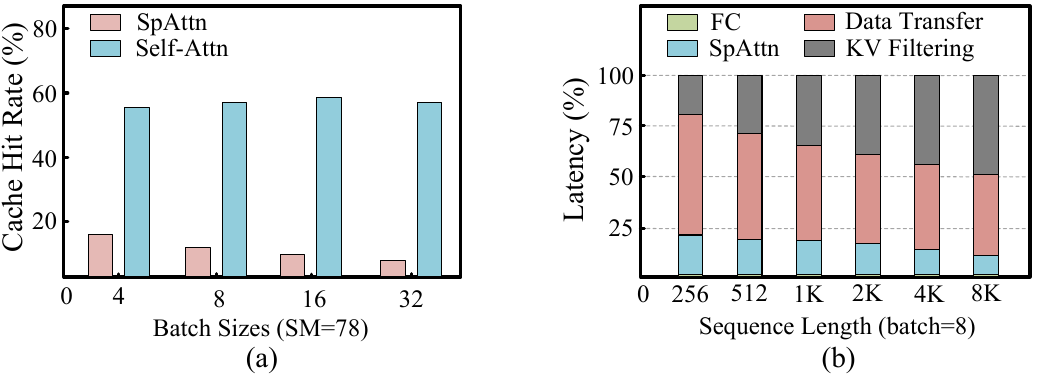}
\vspace{-1em}
\caption{(a) Memory bandwidth saturation analysis as the number of \textit{streaming multiprocessors} (SM) increases, (b) Temporal data locality evaluation by varying batch size while keeping cache capacity fixed}
\label{case1}
\vspace{-1em}
\end{figure}

\section{Motivation Analysis}
\label{motivation_ana}
{ The irregular sampling and low locality of MSDAttn make it fundamentally different from standard self-attention. To quantify its inefficiency on existing platforms, we conduct two case studies.}

\subsection{Case Study\#1---Deformable DETR With GPU}
\textbf{Configurations.} To evaluate the benefits of near-memory processing for DETR, we analyze the open-source \textit{Deformable DETR} (DE-DETR) model~\cite{zhu2021deformable} on an NVIDIA RTX A6000 GPU. Our evaluation focuses on object detection inference using the COCO 2017 dataset~\cite{Lin2014}. To assess the impact of data parallelism, we vary the inference batch size from 1 to 32 with darker colors representing larger batch sizes, where 32 is the standard batch size in~\cite{zhu2021deformable}. Memory bandwidth utilization is measured using the \textit{CUDA Profiling Tools Interface} (CUPTI), while \textit{Nsight Compute} (NCU) is employed to analyze L1 and L2 cache hit rates.

\textbf{Bandwidth Roofline Analysis.} We employ the {bandwidth roofline model} provided by \textit{Nsight Compute} (NCU) to illustrate the theoretical performance limits of the GPU platform. Figure~\ref{cm_ratio}(b) presents the {internal bandwidth} data points for the MSDAttn and FC operators. Our analysis reveals distinct computational characteristics for different components of the model, i.e., MSDAttn exhibits low computational intensity but high memory demands, FC is more compute-intensive. The operational intensity of MSDAttn remains low and constant across batch sizes, as it performs a fixed number of grid-sampling operations and corresponding vector-matrix multiplications. In contrast, FC’s operational intensity increases with batch size due to enhanced data reuse, as all queries in the batch share the same learnable weight matrices. Our findings indicate that the \textit{end-to-end model} (E2EM) remains in the memory-bound region, as its operational intensity is primarily determined by the high proportion of MSDAttn ~\cite{xu2024defa}. As batch size increases, both MSDAttn and the full model approach the system’s theoretical performance limits {since bandwidth limitation}.

\begin{figure}[t]
\centering
\includegraphics[width=8.3cm]{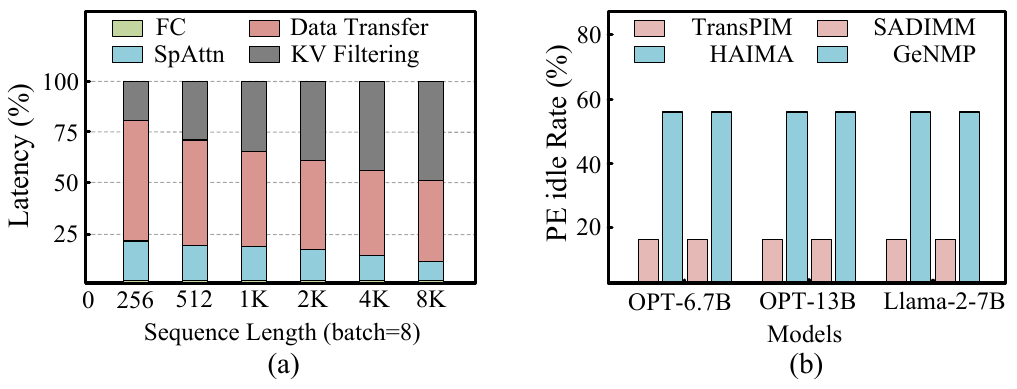}
\vspace{-1em}
\caption{(a) PE idle rate comparison among SOTA NMP attention accelerators, (b) Data reuse rate comparison among SOTA NMP attention accelerators}
\label{case2}
\vspace{-2em}
\end{figure}

\textbf{Compared to Self-Attn.} In Figure~\ref{case1}(a), the bandwidth utilization of both MSDAttn and Self-Attn increases as the batch size grows. This suggests that a larger batch size enhances bandwidth efficiency and ensures fuller utilization of bandwidth. Figure~\ref{case1}(b) presents the cache hit rate of an NVIDIA A6000 GPU when processing MSDAttn and Self-Attn. Compared to Self-Attn, MSDAttn exhibits a significantly lower cache hit rate, which further declines as batch size increases. The strong data locality of Self-Attn arises from the principle that a token is more likely to have strong correlations with its neighboring tokens~\cite{longformer20}. Consequently, in global Self-Attn computation, neighboring token data can be extensively reused. In contrast, MSDAttn relies on highly irregular grid sampling, leading to poor data locality. Furthermore, different batches often involve distinct detection targets, rendering conventional locality-based data reuse strategies ineffective. As illustrated in Figure~\ref{case1}, MSDAttn substantially degrades deformable DETR’s performance, due to excessive memory bandwidth consumption driven by random memory accesses with poor data locality.

\subsection{Case Study\#2---MSDAttn With Modern NMPs}
\textbf{Configurations.} To identify the bottleneck of existing NMP accelerators for MSDAttn, we conduct a case study evaluating modern DETR models on \textit{state-of-the-art} (SOTA) NMP-based attention accelerators. Specifically, we examine three NMP-based accelerators: TransPIM~\cite{Zhou2022}, HAIMA~\cite{Ding2023}, and SADIMM~\cite{sadimm2025}. Since these accelerators are designed for Self-Attn, we introduce an additional module for these platforms to enable MSGS execution during simulation. Our study includes three DETR models, i.e., DE-DETR~\cite{zhu2021deformable}, DN-DETR~\cite{Li_2022_CVPR}, and DINO~\cite{zhang2023dino}. All models are running with object detection inference on the COCO 2017 dataset~\cite{Lin2014}. The data reuse rate is computed as follows: when a query's associated data is accessed for the first time, it is cached by the input buffer following a FIFO replacement policy. If a data block is not reused within the next four queries, it is evicted. The data reuse rate is given by $\frac{NMR-NRE}{NMR}$, where \textit{NMR} is the total number of memory requests, and \textit{NRE} is the number of input buffer replacements. The PE idle rate is calculated as follows: given $N$ processing elements (PEs) and a total execution time $T$ for the NMP accelerator. The execution time and stall time for PE$_i$ are denoted as $t_i$ and $s_i$ ($T = s_i + t_i$), respectively. The average stall ratio across all PEs is given by $\frac{\sum_{i=1}^{N}s_i}{NT}$.

\begin{figure}[t]
\centering
\includegraphics[width=8cm]{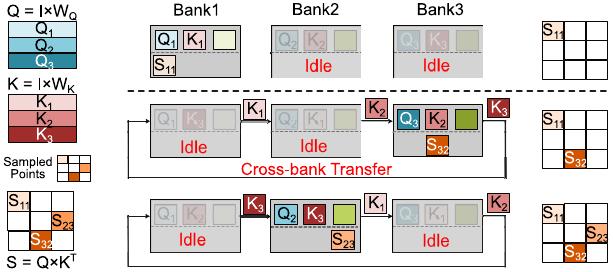}
\vspace{-1em}
\caption{Many PE idle in TransPIM for sampled computing}
\label{imbalance}
\vspace{-1em}
\end{figure}

\textbf{Bottleneck\#1: load imbalance leads to PE idleness.} Figure~\ref{case2}(a) illustrates the PE idle rate for state-of-the-art NMP-based attention accelerators when executing DETR models, revealing that more than 50\% of PEs remain idle. This inefficiency is particularly evident in TransPIM~\cite{Zhou2022}, as shown in Figure~\ref{imbalance}, when applied to MSDAttn. TransPIM adopts a token-based dataflow, where tokens are partitioned and distributed across different memory banks. This approach is highly effective for self-attention, as it ensures high PE parallelism. However, when performing matrix computations with sampling, a significant number of PEs remain inactive. For instance, consider computing attention scores $S = Q \times K^\mathsf{T}$ using sampled elements $S_{11}$, $S_{23}$, and $S_{32}$ from matrix $S$. Although the storage of tokens across banks appears evenly distributed, the sampling operation bypasses most computations. As a result, in each iteration, only the PEs that contain the sampled data remain active, while others remain idle. This severe computational imbalance leads to substantial resource underutilization, limiting the efficiency of NMP-based accelerators when applied to MSDAttn.

\begin{figure*}[t]
\centering
\includegraphics[width=16.5cm]{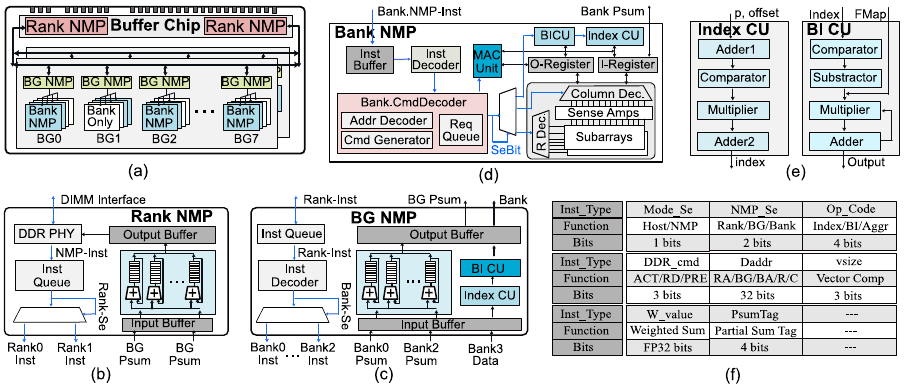}
\vspace{-1em}
\caption{(a) Architecture overview of \textit{DANMP}, (b) Rank-NMP, (c) Bank-group-NMP, (d) Bank-NMP, (e) Details of \textit{index computing unit} (ICU) and \textit{bilinear interpolation computing unit} (BICU), (f) NMP instruction format}
\label{architecture}
\vspace{-1em}
\end{figure*}

\textbf{Bottleneck\#2: poor data reuse leads to bandwidth waste.} Figure~\ref{case2}(b) depicts the data reuse rate of SOTA NMP-based attention accelerators on DETR models, showing a reuse rate of less than 20\%. To understand the root cause of this inefficiency, we analyze TransPIM’s design. TransPIM employs a token-based dataflow to improve data reuse. In the left part of Figure~\ref{imbalance}, the attention score matrix $S$ is computed as $S = Q \times K^\mathsf{T}$. Once $Q_1$ is multiplied by $K_1$, it can be immediately reused to compute $Q_1 \times K_2$, maximizing data locality. However, this reuse strategy fails when accelerating MSDAttn. The primary issue lies in MSGS, which exhibits poor data locality both within a single query and across multiple queries. \textit{Within a single query:} MSGS samples reference points from multiple orientations to capture diverse features, inherently limiting opportunities for data reuse. \textit{Across different queries:} Queries are processed in a random order, with varying detection targets, leading to minimal overlap between adjacent queries. As a direct consequence, a significant portion of memory bandwidth is wasted on repeatedly accessing irregularly sampled data, severely degrading system performance. Therefore, adapting NMP-based self-attention accelerators like TransPIM to MSDAttn is highly inefficient, as the lack of exploitable data locality negates the benefits of dataflow optimization techniques.

\subsection{Key Ideas of DANMP}
\label{key_idea}

%{While recent vision foundation models like LLaVA~\cite{liu2023visual} and Gemini~\cite{team2024gemini} rely on dense attention (e.g., ViT), deformable attention remains essential in tasks with geometric variability, long-range dependencies, or memory constraints---such as object detection~\cite{zhu2021deformable, Li_2022_CVPR, zhang2023dino} and 3D scene understanding~\cite{li2023dfa3d, mao2021voxel, kim2023voxel}. Our goal is not to replace dense vision transformers in general-purpose pipelines, but to improve the efficiency of sparse attention variants like MSDAttn, whose inductive biases and memory efficiency are advantageous in such domains. This focus is particularly relevant for edge AI, embedded vision, and spatially sparse tasks, where dense attention may be unnecessary or inefficient.}

Our first case study reveals that GPU-based MSDAttn suffers from severe memory bandwidth limitations, operating near the roofline. Consequently, leveraging NMP architectures to alleviate these bottlenecks is essential. Our second case study shows that existing NMP-based accelerators are ill-suited for MSDAttn. While current implementations typically offload both FC and self-attention layers, we find that the FC layer is compute-intensive with significant weight reuse. Thus, designing a dedicated NMP accelerator for MSDAttn is urgently needed.

%\textbf{Architecture Designs.} Our second case study observes that existing NMP architectures fail to meet the specific computational and memory access demands of MSDAttn. To address these limitations, we propose a new NMP architecture with three key design components. \textit{Lightweight Near-Memory Integration:} Since the compute-intensive FC layer does not require offloading, we adopt a lightweight integration strategy within memory, ensuring that MSDAttn’s computational needs are met while minimizing area and power consumption. \textit{Uneven Hardware Integration:} We design an unevenly integrated architecture for mapping irregular MSDAttn workloads to significantly reduce PE idleness without introducing cross-bank transfers. \textit{Custom Instruction Set:} We introduce a new instruction set tailored to support our novel architecture design, ensuring efficient execution of MSDAttn operations.

%\textbf{System Designs.} To efficiently support MSDAttn execution on NMP architectures, we introduce two key software-level optimizations. \textit{Clustering-and-Packing (CAP) for Data Reuse:} To increase data reuse, we propose a clustering-and-packing method that groups queries with similar sub-targets, thereby improving data locality and reducing random memory access overhead.
%\textit{Optimized Programming Model \& Execution Flow:} We design a new programming model and execution strategy specifically tailored for \textit{DANMP}, ensuring efficient scheduling and resource utilization in MSDAttn acceleration.

\section{DANMP Architecture Design}
\label{sec:architecture}

\subsection{Overall Architecture}

The overall architecture of \textit{DANMP} is illustrated in Figure~\ref{architecture}(a). It integrates three types of NMP units: near-rank PEs, near bank-group PEs, and near-bank PEs. The buffer chip contains two near-rank PEs responsible for managing I/O data transfers between two ranks. At the bank-group level, each group within a rank is paired with a dedicated PE, enabling near bank-group processing that enhances bandwidth utilization. Unlike prior NMP accelerators such as SADIMM~\cite{sadimm2025} and TransPIM~\cite{Zhou2022}, \textit{DANMP} selectively designates only a subset of banks (50\% based on the measured idle rate in Figure~\ref{case2}(a)) as bank-level NMP units instead of distributing computation across all banks. Each near-bank PE operates in local access mode, meaning it only accesses its corresponding bank (e.g., PE$_i$ exclusively accesses bank$_i$). Furthermore, \textit{DANMP} avoids cross-bank data transfers, a major source of bandwidth inefficiency. Such transfers not only require extra hardware, as in TransPIM~\cite{Zhou2022}, but also incur latency several times higher than local access.

As shown in Figure~\ref{msattention}, MSDAttn consists of two key operations: multi-scale grid sampling and aggregation of extracted features. These rely on three computational primitives—addition, multiplication, and reduction. Integrating \textit{DANMP} therefore requires including these primitives in hardware. A crucial design consideration is the optimal placement of logic units to handle MSDAttn’s varied demands efficiently. \textit{MSGS Execution:} Since MSGS frequently accesses feature maps and sampling points stored in memory banks, executing it with Bank-NMP minimizes latency and maximizes parallelism by keeping data local. \textit{Aggregation Execution:} Aggregation involves extensive reduction operations on interpolated and weighted results. As reduction naturally aligns with DIMM hierarchy, Rank-NMP and BG-NMP efficiently perform this task, leveraging DDR5 structure to optimize data flow and parallelism.

\subsection{Integration Details}
\label{integration_detail}

{\bf Rank-NMP Module.} The Rank-NMP module, shown in Figure~\ref{architecture}(b), integrates near-rank PEs to manage final data aggregation efficiently. In addition to performing standard DDR PHY functions, Rank-NMP serves as a transport layer for NMP control instructions, ensuring low-latency instruction propagation and execution synchronization. {The instruction queue size is limited to 5 because it shares the limited on-die area with NMP logic units. A larger queue can slightly improve packing efficiency but also increases area cost and reduces available compute resources. Through RTL validation, we found that a queue size of 5 achieves the best trade-off between packing benefit and hardware efficiency.} Instructions are multiplexed to specific ranks using the NMP-Select ($\mathtt{NMP\_Se}$) instruction. Rank-NMP aggregates results across all bank-groups. After each bank-group processes its local results (e.g., interpolated features or query partial sums), Rank-NMP performs final data aggregation and reduction. This process accumulates contributions from different feature map levels, forming the final attention-weighted feature vector. For multi-head attention, Rank-NMP also concatenates outputs from different attention heads. Since most arithmetic operations occur at the bank level, Rank-NMP only performs lightweight accumulation, ensuring minimal processing overhead. By finalizing aggregation at the rank level, \textit{DANMP} reduces the data volume sent to the host, improving bandwidth efficiency.

{\bf Bank-NMP Module.} The Bank-NMP module, illustrated in Figure~\ref{architecture}(d), handles MSGS near the bank's sense amplifiers, reducing latency and bandwidth usage. {It includes two primary components (Figure~\ref{architecture}(e)).} The \textit{Index Computation Unit (ICU)} calculates sampled coordinates from scaled reference points, performs boundary checks, and generates memory indices. Microarchitecturally, it comprises a coordinate sampler (Adder1), a comparator-based boundary checker, and an index generator (Multiplier and Adder2). The \textit{Bilinear Interpolation Computational Unit (BICU)} extracts fractional components, computes weights, and performs weighted summation over four neighboring pixels. It consists of a bitwise fraction extractor (Comparator and Subtractor), a multiplier, and a binary adder tree. Positioned near DDR5 sense amplifiers, Bank-NMP PEs perform low-latency memory accesses. Each PE directly computes interpolated features, eliminating raw pixel transfers over the bus. These PEs utilize SIMD-based MAC units for parallel processing. {As the MAC follows conventional architectures like Eyeriss~\cite{chen2016eyeriss}, we omit its internal structure for brevity.} Feature maps are partitioned so each bank holds a spatial tile, enabling bank-level parallelism (details in Section~\ref{sharding}).

{\bf BG-NMP Module.} The BG-NMP module (Figure~\ref{architecture}(c)) acts as a bridge between Rank-NMP and Bank-NMP, aggregating intermediate results while managing bilinear interpolation for non-PE banks. Given the hierarchical nature of DRAM, batch-wise reduction operations are efficiently executed at the bank-group level before forwarding the aggregated data to Rank-NMP for final processing. BG-NMP also coordinates indexing computations, ensuring that near-bank PEs handle hot data within the nearest bank, while bank-group PEs process cold data from all its non-PE banks. Compared to even integration, this approach reduces costly hot data transmission from with-PE bank to non-PE bank, preventing unnecessary movement (cross-bank transfer) of raw pixel data. {To support these functions, each bank-group integrates accumulator, ICU, and BICU units, where the ICU and BICU contain dedicated adder and multiplier arrays capable of performing vector-level MAC operations.} Additionally, BG-NMP balances workloads by distributing simple, localized computations to individual banks while managing cross-bank coordination at the bank-group level. This structure maximizes parallel processing while keeping computations close to the data, ensuring optimal memory efficiency and performance.

%{\bf Figure Clarifications.} To improve clarity, we have explicitly described the components shown in Figure~\ref{architecture}(b)–(d). The {\em Inst Buffer} in Figure~\ref{architecture}(d) receives decoded instructions from the bank-group instruction decoder, which specify whether each operation (e.g., interpolation, accumulation, or reduction) should be executed by local PEs or by the BG-level unit. The {\em SeBit} denotes the row-selection signal within the DRAM array, controlling which wordline is activated for each memory access. The adders shown in Figures~\ref{architecture}(b) and (c) serve as reduction units that accumulate intermediate results collected from multiple banks before forwarding them to higher-level aggregators.

%{{\bf Comparison with Workload-aware NMPs.} Prior NMP accelerators like HAIMA~\cite{Ding2023} and SADIMM~\cite{sadimm2025} employ workload-aware integration by embedding different logic units across the memory hierarchy. HAIMA separates compute units between SRAM and DRAM to handle distinct tasks, while SADIMM performs coarse-grained specialization at the near-bankgroup and near-bank levels. However, both use uniform logic units across all banks, ignoring spatial workload variation. In contrast, \textit{DANMP} adopts a fine-grained, bank-level non-uniform PE integration strategy, placing processing elements in only a subset of banks. To our knowledge, \textit{DANMP} is the first NMP accelerator to support such non-uniform bank-level integration, which is critical for efficiently handling MSDAttn’s highly irregular memory access patterns.}

{{\bf Technical Feasibility of Non-Uniform PEs.} While traditional DRAM (e.g., DDR4/5) employs symmetric bank configurations for manufacturing and interface compatibility, recent research demonstrates the feasibility of heterogeneous designs across banks and subarrays. FIGARO~\cite{wang2020figaro} enables subarray-level partitioning into fast and slow regions using a shared global row buffer. CROW~\cite{hassan2019crow} introduces "regular" and "copy" rows within subarrays, activated independently via additional decoder logic to accelerate hot rows. LISA~\cite{chang2016low} adds lightweight inter-subarray links and identifies fast segments for in-DRAM caching, enabling intra-bank latency heterogeneity. These efforts show that DRAM can be configured at fine granularity, such as banks and subarrays, supporting the non-uniform PE integration strategy proposed in this work.}

\subsection{NMP Instruction Format}
\label{instruct}
{\bf Overview.} {Similar to RecNMP~\cite{recnmp2020} and ReCROSS~\cite{Liu2023},} the \textit{DANMP} accelerator uses a 83-bit compressed instruction format that tightly encodes memory and compute operations. This format was designed to fit within standard memory interface constraints. We refine and elaborate each field of the instruction to better support advanced operations, e.g., indexing, bilinear interpolation, aggregation, weighted sums.

{\bf Instruction Format Breakdown.} The fields and their typical bit widths are shown in Figure~\ref{architecture}(f). \textit{Mode Selection:} The 1-bit $\mathtt{Mode\_Se}$ indicates the memory working on DRAM mode or NMP mode. {In \textbf{DRAM mode}, data remain striped across chips and accessed via host controller. In \textbf{NMP mode}, each bank issues local requests through its embedded PE and executes independently. Feature tiles are remapped into per-bank scratchpads under a separate logical address space, invisible to the host, preserving full DRAM compatibility.} {\textit{NMP Selection:} The 2-bit $\mathtt{NMP\_Se}$ determines whether an instruction is executed by the near-rank PEs, near bank-group PEs, or near-bank PEs. This classification is generated by the host-side memory controller based on operation types.} \textit{Operation Type:} The 4-bit $\mathtt{Op\_Code}$ covers sum/mean ($\mathtt{NMP\_Sum/Mean}$), weighted sum ($\mathtt{NMP\_WSum}$), bilinear interpolation ($\mathtt{NMP\_Interp}$), etc. \textit{DDR Command Flags:} The 3-bit $\mathtt{DDR\_cmd}$ compresses the DRAM commands (Activate, Read, Precharge) needed for the memory access. The memory controller (host) or the near-rank controller sets these flags based on whether the target row is already open, how many column bursts are needed, etc. 
%This compression lets one NMP instruction trigger a sequence of DRAM operations internally, instead of sending each command over the bus.

\textit{Memory Address:} The 32-bit $\mathtt{Daddr}$ supplies the memory location for the operation. In a DRAM setting it can be broken into rank, bank-group, bank, row, and column bits. \textit{Vector Size/Length:} The 3-bit $\mathtt{vsize}$ influences how many DRAM column bursts the rank controller issues. \textit{Weight Field:} The 32-bit $\mathtt{W\_value}$ supplies the scalar weight to multiply the fetched vector by for weighted sum opcodes. \textit{Partial Sum Tag:} The 4-bit $\mathtt{PsumTag}$ identify which accumulation stream this instruction belongs to. This is crucial for aggregation/reduction operations. Only data with the same $\mathtt{PsumTag}$ will be accumulated, ensuring the correctness of the reduction operation.

%If the operation is unweighted, this field can be repurposed. For example, in a bilinear interpolation mode, the 32-bit field could be split into two 16-bit fractional weights (for the X and Y interpolation ratios) instead of a single multiplier.

% This entire flow is controlled by synchronized instructions – for example, a single high-level “compute attention” command might trigger a predefined sequence in the near-rank controller, which in turn coordinates lower-level operations with minimal handshaking. The instruction propagation is hierarchical broadcast: one command from the rank fan-outs into multiple bankgroup commands, which further fan-out to multiple bank operations. Because each level has its own simple controller, the scheduling can be optimized locally (each bankgroup knows to wait for all its banks to finish, etc., without burdening the global controller with those details). This hierarchical control reduces overhead and scales well: adding more banks or groups doesn’t linearly increase complexity at the top.

\section{DANMP System Design}
\label{system}

\subsection{Data Sharding and Mapping}
\label{sharding}
To leverage DDR5’s high bank-level parallelism, feature maps and attention weights are evenly distributed across DDR5’s 32 banks, organized into 8 bank-groups. This partitioning ensures maximum concurrent access by minimizing contention. For instance, a query's sampling points are placed in separate banks to prevent overloading and enable parallel fetching. The four neighboring pixels of a sampling point are stored in the same bank to reduce cross-bank overhead. However, while this tile-based mapping prevents bank conflicts, it introduces load imbalance. Despite equal storage, pixel access probability varies significantly; for example, in portraits, the facial region sees higher sampling, causing uneven workloads.

To address this, we implement a coordinate clustering approach before mapping feature maps to banks. Specifically, query-related coordinates are first computed and clustered to identify frequently sampled regions. {DANMP classifies feature entries by access frequency: the top 50\% most frequently accessed entries are mapped to PE-equipped banks for local computation, while the remaining 50\% are evenly assigned to the banks without PEs.} This frequency-based mapping ensures balanced utilization without overloading PE banks. Since these infrequently accessed data are typically stored across different banks, we use near bank-group PEs to process them, thereby reducing cross-bank data transfers. This optimized mapping strategy incurs minimal overhead, as the coordinate clustering approach operates solely on coordinate data and does not require feature map access or interpolation operations. 
%By balancing computational workload across banks while maintaining high memory access parallelism, \textit{DANMP} achieves efficient and scalable near-memory processing for MSDAttn acceleration.

\subsection{Clustering-and-Packing Algorithm}
\label{capa}
\textbf{Current Challenges in Data Reuse.} In a single DETR query, pixels from various orientations are sampled to capture richer features. However, this process inherently results in low data reuse due to the lack of spatial locality. When executing a batch of queries on a single feature map, data reuse opportunities remain limited for two primary reasons. First, different queries typically correspond to distinct detection targets, which are randomly distributed across the feature map. Second, the arrival order of queries is entirely random, leading to poor temporal locality and further hindering efficient data reuse. Addressing these challenges is critical to minimizing redundant memory accesses and improving system performance.

\textbf{Opportunity Analysis.} Although queries generally focus on different detection targets, many targets can be decomposed into smaller sub-targets, often leading to overlaps across queries. For instance, in portrait detection, detecting both the upper body and face involves extracting common facial features. If queries with shared sub-targets can be identified, spatial locality can be leveraged to enhance data reuse. However, random query order greatly undermines this benefit. For example, after detecting a face, if subsequent queries focus on unrelated objects, the cached facial features may be evicted from the buffer. Consequently, when the system later detects the upper body, it cannot reuse previously computed facial features. To maximize data reuse, queries with shared sub-targets should be grouped to improve temporal locality. This strategic grouping keeps frequently accessed data available for reuse, reducing memory access and improving performance.

\begin{algorithm}
\caption{Clustering-and-Packing}\label{alg:sap}
\begin{algorithmic}[1]
\Require \textit{Q}, \textit{V}, $W^{S}$ are query, FMap and weight matrices, 
$P(x,y)$ is the coordinate set of all reference points
\Ensure Bilinear Interpolation (\textbf{BI}) of sample points $p+\Delta p$
\State $\hat{Q}$ $\leftarrow$ Random selecting 20\% queries from $Q$
\State $\Delta\hat{p}$ = $\hat{Q}\times W^S$
%\State Output $\leftarrow$ \textbf{BI}(V, $\hat{p}+\Delta \hat{p}$)
\State $k\times$ Cluster Centroids $C$ $\leftarrow$ \textbf{K-means}($\hat{p}+\Delta \hat{p}$)
{\State Mapping feature values of $9\times 9$ pixels near $C$ to banks with PE}
\While{$q \in (Q-\hat{Q})$ ($i=0$)} 
\State $\Delta p = q\times W^S$
\State $j^{th}$ Cluster Centroid $C_j$ $\leftarrow$ \textbf{Search}($C$, $p+ \Delta p$)
\State Pack$_j$ = \textbf{Packing}($p+ \Delta p$), $i++$
\EndWhile
\State Output $\leftarrow$ \textbf{BI}(V, from Pack$_0$ to Pack$_k$)
\end{algorithmic}
\end{algorithm}

\textbf{Algorithm Details.} Algorithm~\ref{alg:sap} outlines the CAP method for improving data reuse in bilinear interpolation. {\textit{Random Query Selection (Lines 1-2):} 20\% of queries from the same FMap are randomly selected to compute sampled point offsets $\Delta\hat{p}$. \textit{Clustering (Line 3):} K-means clustering is applied to the sampled points ($\hat{p} + \Delta\hat{p}$), yielding $k$ cluster centroids. A $9\times9$ pixel region is used as the clustering distance metric, identifying frequently accessed (hot) regions. Since only coordinates are involved, this step can be performed prior to FMap storage. The feature values around each cluster center are stored in PE-equipped banks (hot entries), while the rest are placed in non-PE banks (cold entries). \textit{Query Packing (Lines 4-8):} For the remaining 80\% of queries, offsets are computed as $\Delta p = q \times W^S$. Each sampled point ($p + \Delta p$) is mapped to its nearest cluster centroid. Queries targeting the same $9\times9$ region are grouped into packs, enabling shared sub-targets and improved data reuse. \textit{Optimized Bilinear Interpolation (Line 9):} Query packs undergo bilinear interpolation collectively, maximizing reuse and improving efficiency.}

\subsection{Programming Model and Execution Flow}
\label{execution_flow}
{\bf Overview.} At the core of DANMP's programming model, the CPU serves as the system’s central controller, managing global information and orchestrating execution, while NMP units focus solely on executing assigned computations and returning results. Several global factors are critical to ensuring efficiency, including: (1) the distribution of near-bank PEs, (2) the spatial distribution of sampling points, and (3) cluster center information from the CAP algorithm. To illustrate, consider a feature map: the CPU first calculates sampling point locations for 20\% of queries and applies CAP clustering to identify frequently accessed (hotspot) regions. Using this information, the CPU issues instructions to store hotspot-associated feature map segments in banks with PEs, while non-hotspot regions are allocated to banks without PEs. For the remaining 80\% of queries, once the CPU determines sampling point coordinates, it performs query packing to enhance temporal locality, improving data reuse. Finally, based on the spatial distribution of sampling points, the CPU generates instructions for different NMP levels to optimize execution.

\begin{figure}[t]
\centering
\includegraphics[width=8.1cm]{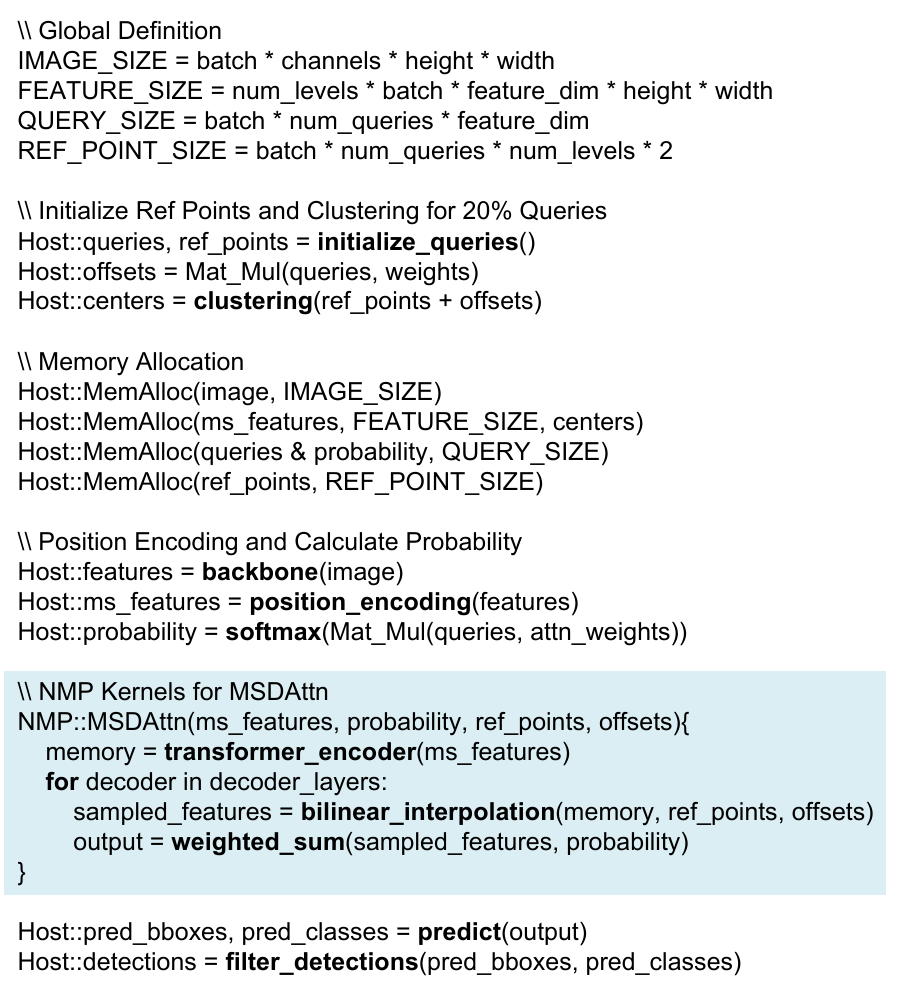}
\vspace{-1em}
\caption{\textit{DANMP} MSDAttn example code}
\label{code}
\vspace{-1.5em}
\end{figure}

{\bf Details of DANMP's Execution Flow.} Figure~\ref{code} presents an example code of host-\textit{DANMP} collaborative processing for DETR inference. The blue region highlights the MSDAttn portion that needs to be offloaded to \textit{DANMP} for execution. We use Figure~\ref{architecture} to elaborate on the detailed process of executing MSDAttn with \textit{DANMP}. The control logic is designed to keep all levels of the hierarchy busy. While one set of sample points is being processed at the banks, the near-rank unit can prepare the next set of instructions, overlapping computation and communication. The execution sequence for one detection query is as follows.

(1) The near-rank NMP receives the NMP instructions of one query, which contains all the informations shown in Figure~\ref{architecture}(f). (2) The near-rank instruction decoder will broadcast the instruction info down to the bank-groups according to the $\mathtt{NMP\_Se}$ field. (3) If $\mathtt{NMP\_Se} = bank$ and the instruction is successfully broadcast to all banks, the banks equipped with PEs will proceed with memory access based on the address specified by $\mathtt{Daddr}$. If $\mathtt{NMP\_Se} = bank-group$, The near-bank-group instruction decoder will collect data from its banks with $\mathtt{Daddr}$. Then the PEs within the bank-groups will generate the bilinear interpolation results locally. (4) Near-bank PEs will collect data from the input buffer in Figure~\ref{architecture}(d). Meanwhile, the near-bank instruction decoder will parse the $\mathtt{Op\_Code}$ field to determine the specific operation to be executed. (5) After the local execution of parallel near-bank PEs, those partial results are sent over a fast intra-bank-group bus to the near-bank-group PE, which adds them together. (6) Each bank-group PE then sends its result (e.g., one interpolated feature vector per sample or a partial sum for the query) up to the near-rank unit. (7) The near-rank unit accumulates all these vectors into the final output for the query's attention computation, and writes it to a designated location or return it to the host.

\section{Evaluation}
\subsection{Experimental Setup}
\label{exp_setup}
{\bf Simulation Framework.} Table~\ref{tab:config} summarizes the core simulation settings used for evaluating the performance of \textit{DANMP}. We conduct a cycle-accurate, full-system simulation of a 32-core architecture with a DDR5-based memory hierarchy. Processor cores and the cache subsystem are modeled using the gem5 simulator~\cite{gem5}, which captures instruction-level behavior, cache interactions, and memory access patterns. To model the memory subsystem, including DDR5 channels and near-memory logic, we extend Ramulator~\cite{ramulator2015} with a custom NMP simulation backend. This extension incorporates cycle-level models for address translation, NMP-specific instruction generation, and detailed execution timing of \textit{DANMP}'s processing elements (PEs). Hardware synthesis of the PEs targets a 40nm CMOS process at 300MHz, matching the internal frequency of DDR5-4800. For area and power estimation of the NMP logic, we utilize Synopsys Design Compiler. Memory access energy and I/O power are derived using CACTI-3DD~\cite{CACTI-3DD} and CACTI-IO~\cite{CACTI-IO}, respectively. 
%All collected energy parameters are integrated into the simulation to provide fine-grained power and energy tracking.

\begin{table}[t]
\centering
\tabcolsep=0.12cm
    \caption{System Parameters and Configurations}
    \vspace{-0.5em}
    \label{tab:config}
    \begin{tabular}{c|c|c|c}  
    \hline\hline
       \multicolumn{4}{c}{\bf Host-side CPU Configurations} \\ \hline
       {Processor} & {32 Cores, 2.5GHz} & {L1 Caches} & {32KB per core}  \\
       \hline
       {L2 Caches} & {1MB per core} & {Shared LLC} & {60MB for 32 cores}  \\
       \hline\hline
       \multicolumn{4}{c}{\bf DRAM Configurations} \\ \hline
       \multicolumn{4}{c}{\makecell{64GB DDR5-4800, 2400MHz, 38.4GB/s per channel, \\4 channels $\times$ 1 DIMM $\times$ 2 Ranks $\times$ 8 bank-groups $\times$ 4 banks, \\ scheduler with a FR-FCFS, 32-entry RD/WR request queue, \\open-page management, Intel Skylake address mapping~\cite{pessl2016drama}}} \\ \hline
       \hline
       \multicolumn{4}{c}{\bf DRAM Timing Parameters (cycles)} \\ \hline
       \multicolumn{4}{c}{\makecell{tRCD = 40, tCL = 40, tRP = 40, tRC = 116, tBL = 8, \\tFAW=32, tRAS = 76, tCCD\_S = 8, tCCD\_L = 12}} \\ \hline
       \hline
       \multicolumn{4}{c}{\bf Latency and Energy Parameters} \\ \hline
       \multicolumn{4}{c}{\makecell{DDR ACT = 2nJ, DDR RD/WR = 4.2pJ/b, Off-chip I/O = 4pJ/b, \\ NMP\_buffer RD/WR = 1 cycle, 50pJ/access, \\ Comparator = 1 cycle, 0.27pJ/Op FP32 adder = 3 cycles, \\0.9pJ/Op, FP32 multiplier = 4 cycles, 2.4pJ/Op }} \\ \hline
       \hline

\end{tabular}
\vspace{-1em}
\end{table}

{\bf Evaluation Methodology.} We benchmark \textit{DANMP} against {six} leading hardware platforms designed for attention mechanisms. These include: (1) A 32-core Intel Xeon Gold 6458Q CPU running at 3.1GHz with a 350W TDP; (2) NVIDIA RTX A6000 GPU equipped with 46GB memory and a 300W TDP, using CUDA v11.6. 
{The GPU baseline employs fused CUDA kernels optimized with shared-memory buffering, warp-level parallelism, and memory coalescing. In addition, we validated our measurements using TensorRT with operator fusion enabled, observing only a marginal ($<$5\%) latency reduction compared to our optimized CUDA version, confirming that the baseline is close to the practical upper bound.}
{(3) DEFA~\cite{xu2024defa}, an ASIC-based deformable attention accelerator leveraging pruning-assisted grid sampling and multi-scale parallel processing to reduce computation overhead;} 
(4) TransPIM~\cite{Zhou2022}, an HBM-based accelerator optimized for self-attention with a token-wise dataflow; 
{(5) HAIMA~\cite{Ding2023}, which combines near-SRAM and near-DRAM to accelerate self-attention using hybrid memory processing; and (6) SADIMM~\cite{sadimm2025}, a DIMM-integrated sparse attention accelerator with dimension-based partitioning to maintain balanced workloads post-pruning.} 
{Since these NMP-based platforms are tailored for traditional self-attention kernels, we extended them with the necessary hardware modules (ICU and BICU units; see Figure~\ref{architecture}(e)) to support multi-scale grid sampling (MSGS) in MSDAttn. All platforms execute inference in FP32 precision. To ensure fair comparison and remove area bias, we report performance using throughput per unit area (GFLOPS/mm$^2$) and energy efficiency using throughput per watt (GFLOPS/W).}

{\bf Workload Configuration.} We evaluate object detection performance across four standard datasets: PASCAL VOC~\cite{everingham2010pascal}, MS COCO~\cite{Lin2014}, KITTI~\cite{kitti2013vision}, and DOTA~\cite{xia2018dota}. Three representative transformer-based detectors are used: DE-DETR~\cite{zhu2021deformable}, DN-DETR~\cite{Li_2022_CVPR}, and DINO~\cite{zhang2023dino}. In our setup, DE-DETR uses 100 detection queries per image, DN-DETR uses 300, and DINO uses 900. {All models are fine-tuned on GPUs using their standard training pipelines to obtain optimized weights, which are then preloaded into DANMP for inference acceleration. No model parameters were modified for DANMP, and its inference accuracy is consistent with GPU and CPU baselines (within ±0.1 mAP on all datasets).}

\begin{figure}[t]
\centering
\includegraphics[width=8.3cm]{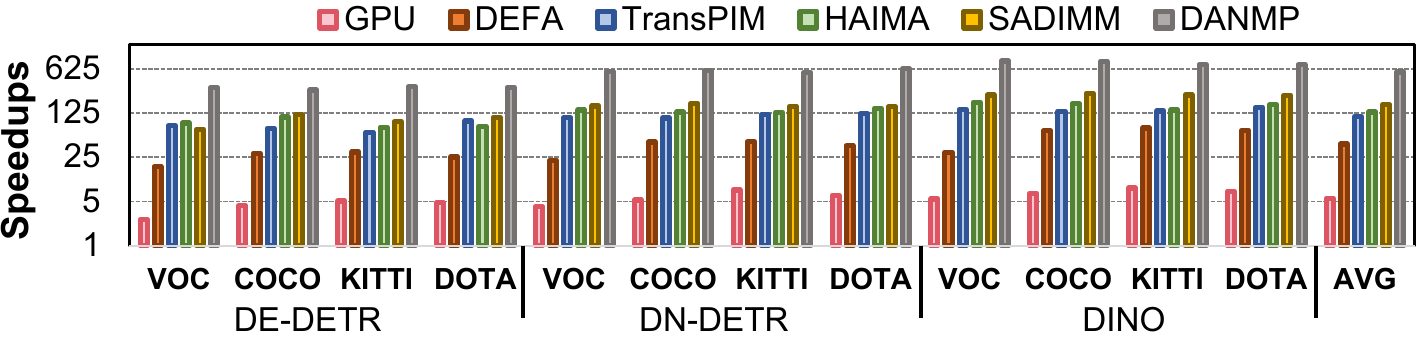}
\vspace{-1em}
\caption{Speedup of \textit{DANMP} compared with GPU, {DEFA}, TransPIM, HAIMA, and SADIMM (normalized to CPU)}
\label{performance}
\vspace{-2em}
\end{figure}

\subsection{Overall Performance}
\label{overall_p}
Figure~\ref{performance} presents the performance of \textit{DANMP} against CPU, GPU, DEFA, TransPIM, HAIMA, and SADIMM.

{\bf Comparison with CPU and GPU.} CPUs and GPUs are the mainstream platforms for accelerating Deformable DETR inference. Across all evaluated datasets and models, \textit{DANMP} delivers an average speedup of 557.31$\times$ over the CPU baseline and 97.43$\times$ over the GPU. This remarkable performance gain stems from three key factors: (1) \textit{DANMP} is specifically optimized—both in hardware and software—for MSDAttn, making it far more suitable for DETR workloads than general-purpose architectures.
(2) Its near-memory architecture minimizes costly host-to-memory data transfers, which typically become a bottleneck in memory-intensive tasks.
(3) By integrating processing units directly at the bank level, \textit{DANMP} unleashes the full potential of DRAM bandwidth, which is essential for the heavy memory access patterns of MSDAttn.

{{\bf Comparison with DEFA.} DEFA is the SOTA ASIC-based accelerator for MSDAttn. Compared to DEFA, DANMP achieves an average speedup of 13.74$\times$, primarily due to two architectural advantages. First, DANMP leverages near-memory parallelism by integrating processing elements directly adjacent to DRAM banks. This design eliminates frequent off-chip memory transfers and enables high-throughput, low-latency data access. Second, DANMP adopts a Clustering-and-Packing (CAP) algorithm that identifies and groups redundant queries across consecutive frames, significantly reducing memory access overhead in streaming workloads.}

{\bf Comparison with TransPIM and HAIMA.} TransPIM and HAIMA are state-of-the-art accelerators optimized for standard self-attention. Nevertheless, \textit{DANMP} surpasses them by 5.17$\times$ and 4.26$\times$ on average, respectively. The reasons for this advantage include: (1) TransPIM and HAIMA target dense matrix multiplications typical of self-attention, whereas MSDAttn involves irregular and sparse operations that GEMM units handle poorly. \textit{DANMP} addresses this with specialized PEs that execute MSGS directly in memory. (2) Their dataflows and schedulers are designed for regular patterns, making them less effective for MSDAttn’s complexity. In contrast, \textit{DANMP} introduces a custom NMP dataflow tailored to MSDAttn's irregular behavior, improving runtime efficiency. (3) Unlike MSDAttn, self-attention workloads are regular, so prior accelerators overlook dynamic load balancing. In MSGS, severe load imbalance can leave some banks with minimal data for their PEs, causing idle cycles after local computations. \textit{DANMP} addresses this with non-uniform hardware integration, aligning resources with workload irregularity to reduce PE idling. Moreover, relocating cold data within a bank-group requires only a read from the source bank, significantly lowering overhead compared to cross-bank transfers. {While DANMP employs fewer PEs than TransPIM, its throughput is higher because both architectures perform the same computation, making latency the key factor. The latency is dominated by the most loaded bank–PE pair. DANMP’s query grouping greatly improves data reuse among nearby queries, reducing redundant data movement and PE idle time. As a result, DANMP achieves lower execution latency for the same workload, delivering higher effective throughput despite using fewer PEs.}

\begin{figure}[t]
\centering
\includegraphics[width=8.3cm]{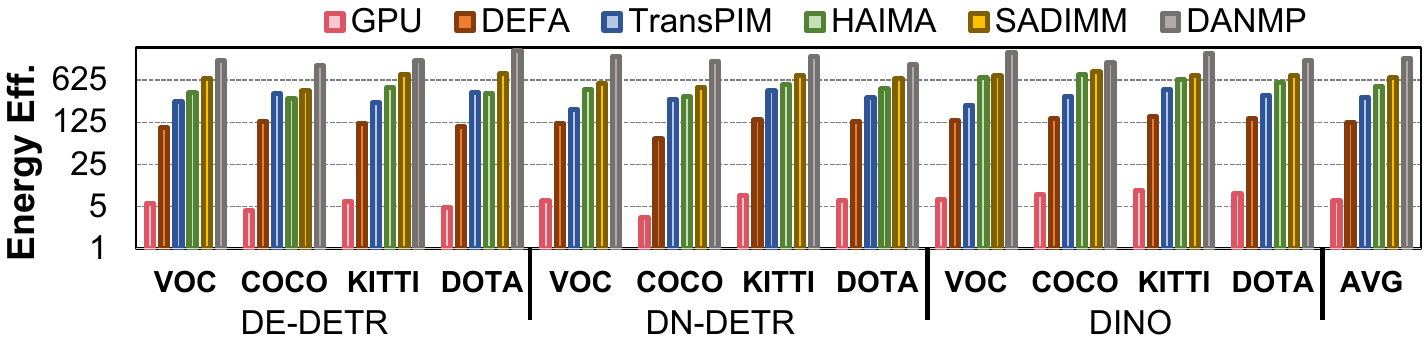}
\vspace{-1em}
\caption{Energy efficiency of \textit{DANMP} compared with GPU, {DEFA}, TransPIM, HAIMA, and SADIMM}
\label{energy}
\vspace{-1em}
\end{figure}

{\bf Comparison with SADIMM.} Compared to SADIMM, a leading sparse-attention accelerator, \textit{DANMP} achieves a 3.43$\times$ average speedup. The improvements arise from several innovations: (1) While both adopt efficient memory-side execution, \textit{DANMP} offers a dataflow better suited for MSDAttn. (2) SADIMM introduces a dimension-wise data partitioning strategy to balance load across PEs. However, this method proves insufficient for MSDAttn’s more irregular computation, often resulting in idle compute resources. \textit{DANMP} addresses this through its irregular hardware mapping, aligning hot (cold) entries to bank-level (bank-group level) PEs to ensure balanced execution. (3) SADIMM enhances reuse through diagonal locality in sparse attention~\cite{asadi2024}. Unfortunately, MSDAttn lacks such regularity. \textit{DANMP} compensates by applying the Clustering-and-Packing (CAP) algorithm, which groups queries with shared sub-targets to boost both spatial and temporal data reuse.

{\bf Energy Efficiency Comparison.} Figure~\ref{energy} shows energy efficiency results. {Compared to CPU, GPU, and DEFA platforms, \textit{DANMP} achieves 1437.82$\times$, 227.21$\times$, and 11.63$\times$ improvements, respectively.} {
For completeness, DRAM refresh, DIMM I/O termination, and NMP instruction decoding overheads are also included in the revised estimation.
Their combined contribution accounts for less than 8\% of total system energy, leading to an adjusted overall efficiency of 208.47× over GPU baseline.
} {Compared to the GPU, DANMP achieves a higher improvement in energy efficiency than in performance. This disparity stems from DANMP’s ability to avoid costly off-chip memory accesses, the dominant source of energy consumption. In contrast, GPUs suffer from substantial energy overhead due to frequent data movement and always-on compute resources.} Relative to other NMP accelerators, \textit{DANMP} achieves 4.41$\times$, 2.85$\times$, and 2.37$\times$ better energy efficiency than TransPIM, HAIMA, and SADIMM, respectively. These gains can be attributed to: (1) \textit{DANMP}’s non-uniform integration strategy ensures that near-bank PEs are only deployed where needed, reducing energy waste from idle units. In contrast, platforms with uniform integration must reallocate tasks through costly cross-bank communication. (2) The clustering-and-packing algorithm also significantly increases data reuse, reducing redundant memory accesses and thereby lowering energy per operation.

\begin{figure}[t]
\centering
\includegraphics[width=8.3cm]{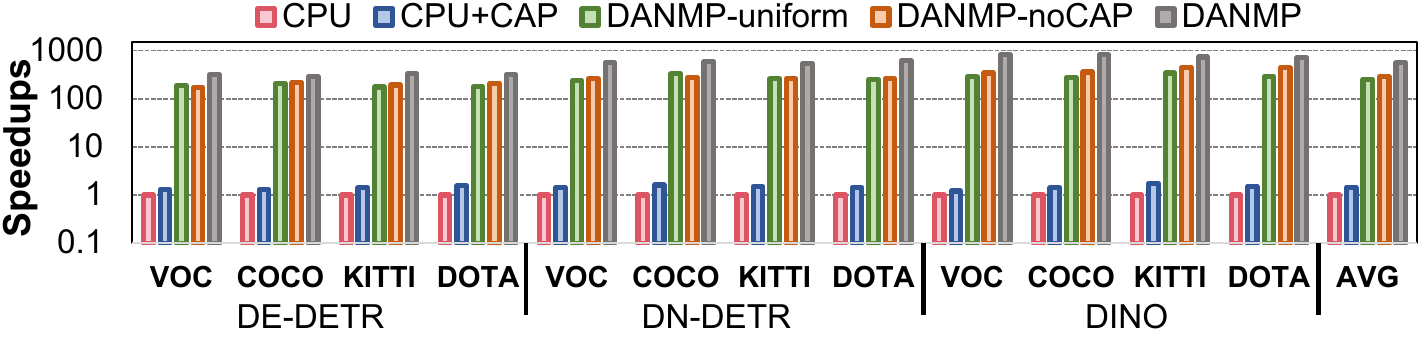}
\vspace{-1em}
\caption{Performance comparison of extra configurations}
\label{softonly}
%\vspace{-2em}
\end{figure}

\subsection{Hardware and Software Efficiency}
\label{hard_soft}
To better understand the individual contributions of \textit{DANMP}'s hardware and software components, we conducted experiments across three supplementary platforms: (1) CPU+CAP, where the CAP algorithm is applied during DETR query inference on the CPU to improve data locality; (2) DANMP-uniform, where identical bank-level PEs are integrated into all banks in DANMP; and (3) DANMP-noCAP, which runs \textit{DANMP} without CAP to isolate its software impact. {The results are shown in Figure~\ref{softonly}.}

{\bf Software Effectiveness.} {Compared to the CPU baseline, integrating CAP yields a 1.45× speedup, demonstrating its effectiveness. Although it introduces clustering and packing overhead, CAP significantly improves data reuse and reduces redundant memory accesses.} This gain arises from two key benefits. First, CAP organizes queries sharing common sub-targets, avoiding repeated computations on overlapping regions. Second, by grouping spatially correlated queries, CAP improves temporal locality, minimizing redundant memory accesses and mitigating random access overhead. These findings support the insight that combining multiple ML strategies is often essential for improving system efficiency in practice.

{\bf Hardware Effectiveness.} Compared to DANMP-uniform, DANMP achieves a 2.21× performance boost. Although CAP improves data reuse in DANMP-uniform, its uniform hardware configuration cannot adapt to the irregular access patterns of MSDAttn, causing significant PE underutilization. To avoid idling, data must be transferred from other banks via the MC, which is far less efficient than local bank-group reads. Moreover, DANMP-uniform provides no more compute resources than \textit{DANMP}, as \textit{DANMP} reallocates some PEs to the bank-group level for better load balancing. Even without CAP optimization, DANMP-noCAP still surpasses the CPU baseline by a large margin, achieving a 283.63× speedup. This improvement mainly results from the NMP-based architecture, which minimizes off-chip transfers, and the near-bank computing scheme that significantly increases available bandwidth. Although omitting CAP introduces some bandwidth inefficiencies, \textit{DANMP} still maintains a substantial advantage, underscoring the potential of NMP-based architectures for specialized DETR acceleration.

\begin{figure}[t]
\centering
\subfloat[]{
\begin{minipage}[t]{0.489\linewidth}
\centering
\includegraphics[width=1.5in]{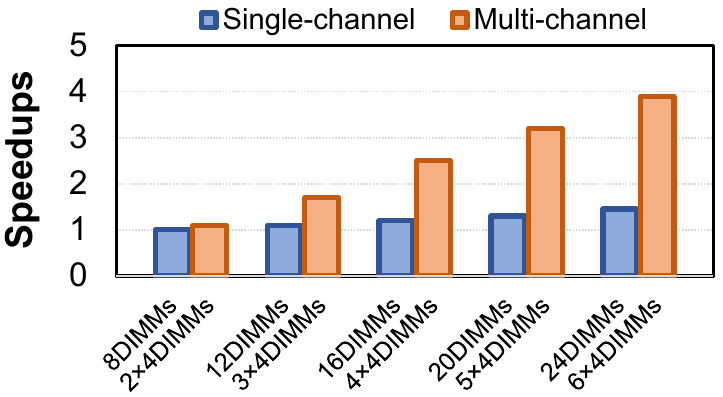}
\end{minipage}
}%
\subfloat[]{
\begin{minipage}[t]{0.489\linewidth}
\centering
\includegraphics[width=1.5in]{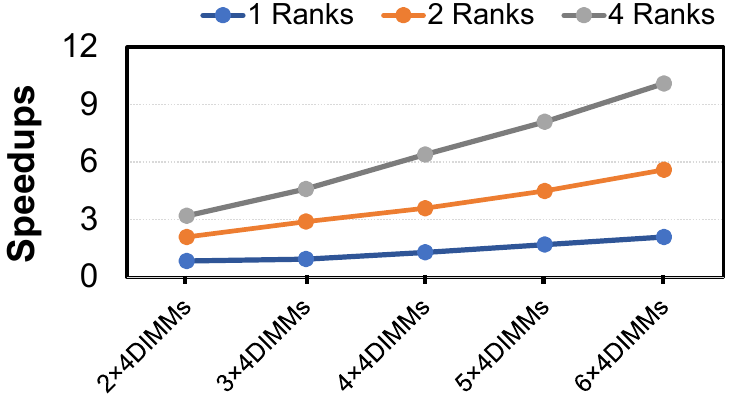}
\end{minipage}
}%
\centering
\vspace{-1em}
\caption{(a) Scalability of \textit{DANMP} on DE-DETR as the number of DIMMs increases, and (b) Performance of \textit{DANMP} with different number of ranks}
\vspace{-1em}
\label{hard_scalability}
%\vspace{-1em}
\end{figure}

\subsection{Scalability}
\label{scala_exp}
{\bf Impact of DIMM and Rank Configuration.} Figure~\ref{hard_scalability}(a) shows \textit{DANMP} performance scaling on MSDAttn workloads {under varying DIMM and memory channel counts.} Results reveal that simply increasing DIMMs on a single memory channel yields limited improvement. Although more DIMMs provide higher internal bandwidth, control signals remain constrained by the single channel, making it the bottleneck. In contrast, adding DIMMs across multiple channels significantly boosts throughput by removing this constraint and enabling parallel C/A paths. Figure~\ref{hard_scalability}(b) further examines performance changes with rank count per DIMM. While adding ranks enhances bandwidth at the rank level, gains are sublinear due to shared bus connections. Nonetheless, performance improves steadily—for example, a 4-rank configuration achieves a 1.63× speedup over 2 ranks, demonstrating that \textit{DANMP} effectively exploits bandwidth hierarchy across ranks.

\begin{figure}[t]
\centering
\includegraphics[width=8.3cm]{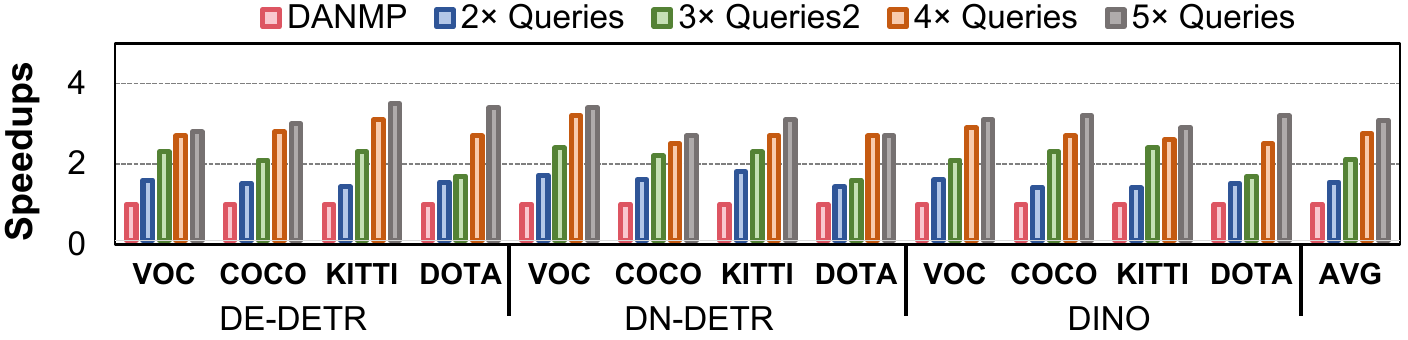}
\vspace{-1em}
\caption{The speedups with various number of queries}
\label{query_scala}
%\vspace{-2em}
\end{figure}

{\bf Impact of Query Volume.} We simulate expanded workloads by proportionally increasing the number of queries and feature map dimensions. For instance, the "2× Queries" setup doubles the query and feature map sizes compared to the baseline. As shown in Figure~\ref{query_scala}, \textit{DANMP}'s performance gain over the CPU baseline continues to grow with the number of queries. This trend highlights \textit{DANMP}’s strong scalability in large-scale inference tasks. First, the use of near-memory computing minimizes the off-chip random memory access overhead. This advantage becomes more pronounced as the number of queries increases. Second, the CAP algorithm enhances temporal and spatial data reuse by clustering queries that target overlapping regions. As the number of queries increases, the probability of shared sub-targets rises, leading to even higher reuse efficiency.

\begin{figure}[t]
\centering
\subfloat[]{
\begin{minipage}[t]{0.489\linewidth}
\centering
\includegraphics[width=1.5in]{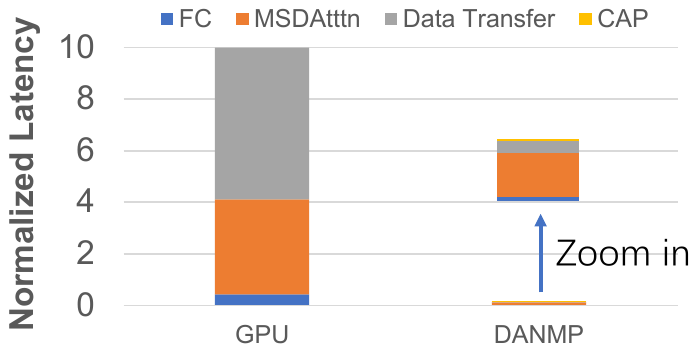}
\end{minipage}
}%
\subfloat[]{
\begin{minipage}[t]{0.489\linewidth}
\centering
\includegraphics[width=1.5in]{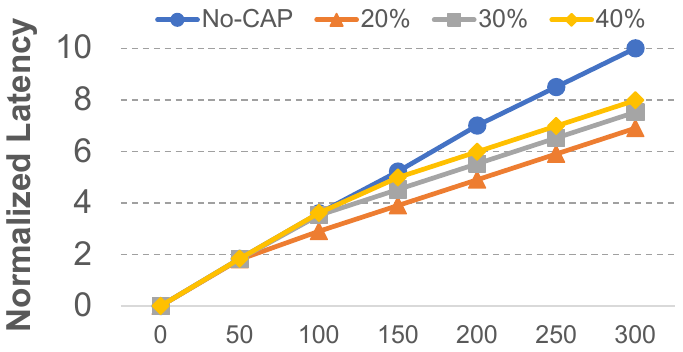}
\end{minipage}
}%
\centering
\vspace{-0.5em}
\caption{(a) Latency breakdown of a DE-DETR transformer block (COCO, 100 queries per FMap, batch size 8), and (b) Impact of CAP clustering ratio on normalized inference latency (CPU, 300 queries per FMap, batch size 8)}
\vspace{-1em}
\label{latency_break}
%\vspace{-1em}
\end{figure}

{{\bf Impact of CAP Sampling Rate.} Figure~\ref{latency_break}(b) presents the experimental results of the impact of CAP clustering ratio on normalized inference latency. {The x-axis represents cumulative query progress (50, 100, 150, 200, 250, and 300 queries), and the y-axis shows the corresponding latency. The clustering-and-packing overhead is already included. Since it is performed only once for 300 queries, and the latency of a single query is higher than the packing overhead..} Compared to the no-CAP baseline, CAP with various sampling ratios consistently achieves lower latency, demonstrating its effectiveness. The lowest latency is observed when 20\% of the queries are selected for coordinate clustering, as the remaining 80\% benefit from improved data reuse enabled by CAP. A smaller ratio, such as 10\%, is not chosen because it yields limited benefits on feature maps with only 100 queries. We set 20\% as the default configuration because it achieves the best trade-off between clustering effectiveness and minimal inference latency.}

\subsection{{Latency/}Energy/Area Breakdown}
\label{lea_break}
{\bf Latency Breakdown.} Figure~\ref{latency_break}(a) shows the latency breakdown of a transformer block in DE-DETR on the COCO dataset, with 100 queries per feature map and a batch size of 8. On GPU, MSDAttn accounts for 36.4\% of total latency, while data movement dominates with 59.2\%, reflecting the low cache hit rate in Figure~\ref{case1}(b). The FC layer contributes less than 5\%, as its regular access pattern overlaps effectively with data transfers, consistent with prior studies~\cite{kao2023flat, gilbert2024looptree, lee2024infinigen}. {In \textit{DANMP}, CAP, FC, data movement, and MSDAttn contribute 1.3\%, 8.6\%, 27.4\%, and 62.7\% of latency, respectively, indicating minimal CAP overhead.} Although the FC layer runs on the CPU, its latency remains low due to overlap with data transfers. {Unlike the GPU, where FC execution is delayed by the long MSDAttn stage due to data dependencies, DANMP’s faster NMP-based MSDAttn shortens this dependency chain, resulting in significantly less FC stall time and reducing the overall latency.} Despite being an NMP accelerator, DANMP incurs 27\% data movement overhead, primarily from transferring weight matrices to the CPU.

{\bf Energy/Area Breakdown.} We present the area and power consumption of SADIMM and \textit{DANMP} in Table~\ref{tab:breakdown}. {In 40nm technology, \textit{DANMP} shows a minimal rank PE area per buffer chip of 0.42 $mm^2$, a notable reduction from SADIMM’s 0.73 $mm^2$, mainly because the rank PE in \textit{DANMP} only handles aggregation and omits softmax. \textit{DANMP} exhibits a bank-group PE area per DRAM chip of 0.57 $mm^2$, slightly larger than SADIMM’s 0.39 $mm^2$, since BG-level PEs in \textit{DANMP} share part of the workload originally handled by bank-level PEs. \textit{DANMP} also shows a minimal bank PE area per DRAM chip of 1.03 $mm^2$, a significant reduction from SADIMM’s 2.29 $mm^2$. Furthermore, \textit{DANMP} consumes 259 $mW$, 31.6\% lower than SADIMM’s 335 $mW$.}

\begin{table}[t]
\centering
\tabcolsep=0.05cm
    \caption{Power and area breakdown of \textit{DANMP}}
    \vspace{-1em}
    \label{tab:breakdown}
    \begin{tabular}{|c|c|c|c|c|}  
    \cline{1-5}
    
       {\bf Name} & {\bf Component}& {\bf PE type} & {\bf Area (mm$^2$) } & {\bf Power (mW)}\\
       \hline \hline

       \multirow{3}*{SADIMM} & {Rank PEs} & {adder\&soft.} & {0.73} & {82} \\
       \cline{2-5}
       {} & {BG PEs} & {adder} & {0.39} & {37} \\
       \cline{2-5}
       {} & {Banks PEs} & {multiplier} & {2.29} & {216} \\
       \hline
       \multirow{3}*{DANMP} & {Rank PEs} & {MAC Unit} & {0.42} & {51} \\
       \cline{2-5}
       {} & {BG PEs} & {ICU\&BICU} & {0.57} & {71} \\
       \cline{2-5}
       {} & {Bank PEs} & {ICU\&BICU} & {1.03} & {137} \\
       \cline{1-5}
    
\end{tabular}
%\vspace{-1.5em}
\end{table}

\subsection{Discussion}
\label{discussion}

{
{\bf Support Other Applications.} We discuss how \textit{DANMP} benefits other memory-bound workloads. A prime example is graph processing, known for its irregularity: high-degree nodes require heavy computation, while low-degree ones are lightweight. Uniform assignment causes runtime inefficiencies. \textit{DANMP}’s heterogeneous integration addresses this by allocating heavy, high-degree nodes to bank-level PEs and lighter ones to bank-group PEs, enhancing efficiency. Similarly, in databases, frequently accessed fields (e.g., primary keys) can be routed to bank-level PEs, leaving less active fields to bank-group units. While not aiming for a universal accelerator, our insights—such as clustering-based reuse and non-uniform PE integration—can guide future NMP architectures for graph analytics, sparse vision, and other irregular applications.}

\section{Related Works}
{\bf Traditional Attention Accelerators.} Early efforts to accelerate Transformer-based models largely leveraged GPUs due to their high parallel computing capabilities~\cite{Vaswani17, bert18, longformer20, Flashatten2022, zhu2021deformable, Li_2022_CVPR, zhang2023dino, Aken19, RoBERTa2019, Arnab_2021_ICCV}. GPU-focused approaches~\cite{longformer20, Flashatten2022, Zaheer20, Performer2020, Linformer2020, wu2020visual, Zaheer20} aim to reduce memory consumption and computational complexity in attention modules. To further alleviate compute and memory overheads, co-designed accelerators have been developed using FPGAs~\cite{Zhang21, bai2024dac, Li20, Ye2023} and ASICs~\cite{Ham20, Lu21, Qu22, xu2024defa, Qin2023, You2023}. These platforms provide better support for custom scheduling and attention-specific logic. However, their performance remains limited by frequent random memory accesses and the increasing cost of off-chip memory transfers, with bandwidth quickly becoming the system bottleneck~\cite{Xie2021, pim-mmu2024}.

{\bf Near-Memory Attention Acceleration.} Near-memory processing (NMP) architectures offer a promising alternative by enabling in-memory data processing and reducing off-chip traffic. These architectures have shown notable improvements in bandwidth utilization and throughput across memory-intensive workloads such as non-Transformer deep learning~\cite{Gao2017, ReGNN2022, Xu2024, Kal2021, Park2021, Liu2023, Angizi2018}, graph processing~\cite{Talati2022, Dai2022, Zhang2018}, and attention models~\cite{li2022cpsaa, Zhou2022, Ding2023, asadi2024, sadimm2025, Yang20, Kang21, Laguna2021, Zhou2021, sprint2022, Li2023, Sridharan2023, Zheng2023}. While NMP designs excel at minimizing memory access latency, many still suffer from inefficient PE utilization due to workload imbalance and limited data reuse opportunities. These shortcomings result in increased scheduling overhead and suboptimal resource use. To address these issues, our work proposes a heterogeneous NMP design with non-uniform integration of compute units and introduces a collaborative optimization framework between the host and memory modules.

\section{Conclusion}
This paper targets the memory-bound MSDAttn mechanism. As existing NMP accelerators struggle with utilization and efficiency, we propose \textit{DANMP}, a hardware–software co-designed solution. \textit{DANMP} employs selective PE integration, placing compute units only in designated banks to match MSDAttn’s irregularity and reduce cross-bank communication. In software, the CAP method ensures efficient execution via a tailored dataflow. Experiments demonstrate that \textit{DANMP} significantly outperforms state-of-the-art NMP accelerators in both performance and energy efficiency.

\bibliographystyle{ACM-Reference-Format}
\bibliography{sample-base}

@ARTICLE{resqm2020,
  author={Li, Huize and Jin, Hai and Zheng, Long and Liao, Xiaofei},
  journal={IEEE Transactions on Computer-Aided Design of Integrated Circuits and Systems}, 
  title={ReSQM: Accelerating Database Operations Using ReRAM-Based Content Addressable Memory}, 
  year={2020},
  volume={39},
  number={11},
  pages={4030-4041},
  doi={10.1109/TCAD.2020.3012860}}

@ARTICLE{cpsaa2024,
  author={Li, Huize and Jin, Hai and Zheng, Long and Liao, Xiaofei and Huang, Yu and Liu, Cong and Xu, Jiahong and Duan, Zhuohui and Chen, Dan and Gui, Chuangyi},
  journal={IEEE Transactions on Computer-Aided Design of Integrated Circuits and Systems}, 
  title={CPSAA: Accelerating Sparse Attention Using Crossbar-Based Processing-In-Memory Architecture}, 
  year={2024},
  volume={43},
  number={6},
  pages={1741-1754},
  doi={10.1109/TCAD.2023.3344524}}

@inproceedings{resma2022,
author = {Li, Huize and Jin, Hai and Zheng, Long and Huang, Yu and Liao, Xiaofei and Duan, Zhuohui and Chen, Dan and Gui, Chuangyi},
title = {ReSMA: accelerating approximate string matching using ReRAM-based content addressable memory},
year = {2022},
isbn = {9781450391429},
publisher = {Association for Computing Machinery},
address = {New York, NY, USA},
url = {https://doi.org/10.1145/3489517.3530559},
doi = {10.1145/3489517.3530559},
booktitle = {Proceedings of the 59th ACM/IEEE Design Automation Conference},
pages = {991–996},
numpages = {6},
location = {San Francisco, California},
series = {DAC '22}
}

@ARTICLE{highp2025,
  author={Chen, Dan and Li, Huize and Lan, Huiying and Li, Zhaoying and Yao, Pengcheng and Mitra, Tulika},
  journal={IEEE Transactions on Computers}, 
  title={HighP: In-Memory Acceleration of SpGEMM With High Bank-Level Parallelism}, 
  year={2026},
  volume={75},
  number={1},
  pages={139-151},
  doi={10.1109/TC.2025.3624920}}

@ARTICLE{splim2025,
  author={Li, Huize and Chen, Dan and Mitra, Tulika},
  journal={IEEE Transactions on Computer-Aided Design of Integrated Circuits and Systems}, 
  title={SPLIM: Bridging the Gap Between Unstructured SpGEMM and Structured In-Situ Computing}, 
  year={2025},
  volume={44},
  number={6},
  pages={2412-2423},
  doi={10.1109/TCAD.2024.3522882}}

@article{parashar2017scnn,
  title={SCNN: An accelerator for compressed-sparse convolutional neural networks},
  author={Parashar, Angshuman and Rhu, Minsoo and Mukkara, Anurag and Puglielli, Antonio and Venkatesan, Rangharajan and Khailany, Brucek and Emer, Joel and Keckler, Stephen W and Dally, William J},
  journal={ACM SIGARCH computer architecture news},
  volume={45},
  number={2},
  pages={27--40},
  year={2017},
  publisher={ACM New York, NY, USA}
}

@article{liang2020engn,
  title={EnGN: A high-throughput and energy-efficient accelerator for large graph neural networks},
  author={Liang, Shengwen and Wang, Ying and Liu, Cheng and He, Lei and Li, Huawei and Xu, Dawen and Li, Xiaowei},
  journal={IEEE Transactions on Computers},
  volume={70},
  number={9},
  pages={1511--1525},
  year={2020},
  publisher={IEEE}
}

@inproceedings{geng2020awb,
  title={AWB-GCN: A graph convolutional network accelerator with runtime workload rebalancing},
  author={Geng, Tong and Li, Ang and Shi, Runbin and Wu, Chunshu and Wang, Tianqi and Li, Yanfei and Haghi, Pouya and Tumeo, Antonino and Che, Shuai and Reinhardt, Steve and others},
  booktitle={2020 53rd Annual IEEE/ACM International Symposium on Microarchitecture (MICRO)},
  pages={922--936},
  year={2020},
  organization={IEEE}
}

@inproceedings{gudaparthi2022candles,
  title={CANDLES: Channel-aware novel dataflow-microarchitecture co-design for low energy sparse neural network acceleration},
  author={Gudaparthi, Sumanth and Singh, Sarabjeet and Narayanan, Surya and Balasubramonian, Rajeev and Sathe, Visvesh},
  booktitle={2022 IEEE International Symposium on high-performance computer architecture (HPCA)},
  pages={876--891},
  year={2022},
  organization={IEEE}
}

@article{gem5,
author = {Binkert, Nathan and Beckmann, Bradford and Black, Gabriel and Reinhardt, Steven K. and Saidi, Ali and Basu, Arkaprava and Hestness, Joel and Hower, Derek R. and Krishna, Tushar and Sardashti, Somayeh and Sen, Rathijit and Sewell, Korey and Shoaib, Muhammad and Vaish, Nilay and Hill, Mark D. and Wood, David A.},
title = {The gem5 simulator},
year = {2011},
issue_date = {May 2011},
publisher = {Association for Computing Machinery},
address = {New York, NY, USA},
volume = {39},
number = {2},
issn = {0163-5964},
url = {https://doi.org/10.1145/2024716.2024718},
doi = {10.1145/2024716.2024718},
journal = {SIGARCH Comput. Archit. News},
month = aug,
pages = {1–7},
numpages = {7}
}

@INPROCEEDINGS{pim-mmu2024,
  author={Lee, Dongjae and Hyun, Bongjoon and Kim, Taehun and Rhu, Minsoo},
  booktitle={2024 57th IEEE/ACM International Symposium on Microarchitecture (MICRO)}, 
  title={{PIM-MMU: A Memory Management Unit for Accelerating Data Transfers in Commercial PIM Systems}}, 
  year={2024},
  volume={},
  number={},
  pages={627-642},
  doi={10.1109/MICRO61859.2024.00053}}

@INPROCEEDINGS{micro2024ham,
  author={Ham, Hyungkyu and Hong, Jeongmin and Park, Geonwoo and Shin, Yunseon and Woo, Okkyun and Yang, Wonhyuk and Bae, Jinhoon and Park, Eunhyeok and Sung, Hyojin and Lim, Euicheol and Kim, Gwangsun},
  booktitle={2024 57th IEEE/ACM International Symposium on Microarchitecture (MICRO)}, 
  title={{Low-Overhead General-Purpose Near-Data Processing in CXL Memory Expanders}}, 
  year={2024},
  volume={},
  number={},
  pages={594-611},
  doi={10.1109/MICRO61859.2024.00051}}

@INPROCEEDINGS{micro2024li,
  author={Li, Yiwei and Tian, Boyu and Ren, Yi and Gao, Mingyu},
  booktitle={2024 57th IEEE/ACM International Symposium on Microarchitecture (MICRO)}, 
  title={{Stream-Based Data Placement for Near-Data Processing with Extended Memory}}, 
  year={2024},
  volume={},
  number={},
  pages={1648-1662},
  doi={10.1109/MICRO61859.2024.00120}}

@INPROCEEDINGS{CACTI-3DD,
  author={Chen, Ke and Li, Sheng and Muralimanohar, Naveen and Ahn, Jung Ho and Brockman, Jay B. and Jouppi, Norman P.},
  booktitle={2012 Design, Automation \& Test in Europe Conference \& Exhibition (DATE)}, 
  title={{CACTI-3DD: Architecture-level modeling for 3D die-stacked DRAM main memory}}, 
  year={2012},
  volume={},
  number={},
  pages={33-38},
  doi={10.1109/DATE.2012.6176428}}

@inproceedings{CACTI-IO,
author = {Jouppi, Norman P. and Kahng, Andrew B. and Muralimanohar, Naveen and Srinivas, Vaishnav},
title = {{CACTI-IO: CACTI with off-chip power-area-timing models}},
year = {2012},
isbn = {9781450315739},
publisher = {Association for Computing Machinery},
address = {New York, NY, USA},
url = {https://doi.org/10.1145/2429384.2429446},
doi = {10.1145/2429384.2429446},
booktitle = {Proceedings of the International Conference on Computer-Aided Design},
pages = {294–301},
numpages = {8},
location = {San Jose, California},
series = {ICCAD '12}
}

@inproceedings{pessl2016drama,
  title={$\{$DRAMA$\}$: Exploiting $\{$DRAM$\}$ addressing for $\{$Cross-CPU$\}$ attacks},
  author={Pessl, Peter and Gruss, Daniel and Maurice, Cl{\'e}mentine and Schwarz, Michael and Mangard, Stefan},
  booktitle={25th USENIX security symposium (USENIX security 16)},
  pages={565--581},
  year={2016}
}

@inproceedings{xia2018dota,
  title={DOTA: A large-scale dataset for object detection in aerial images},
  author={Xia, Gui-Song and Bai, Xiang and Ding, Jian and Zhu, Zhen and Belongie, Serge and Luo, Jiebo and Datcu, Mihai and Pelillo, Marcello and Zhang, Liangpei},
  booktitle={Proceedings of the IEEE conference on computer vision and pattern recognition},
  pages={3974--3983},
  year={2018}
}

@article{everingham2010pascal,
  title={The pascal visual object classes (voc) challenge},
  author={Everingham, Mark and Van Gool, Luc and Williams, Christopher KI and Winn, John and Zisserman, Andrew},
  journal={International journal of computer vision},
  volume={88},
  pages={303--338},
  year={2010},
  publisher={Springer}
}

@InProceedings{Lin2014,
author="Lin, Tsung-Yi
and Maire, Michael
and Belongie, Serge
and Hays, James
and Perona, Pietro
and Ramanan, Deva
and Doll{\'a}r, Piotr
and Zitnick, C. Lawrence",
title="{Microsoft COCO: Common Objects in Context}",
booktitle="Computer Vision -- ECCV 2014",
year="2014",
publisher="Springer International Publishing",
address="Cham",
pages="740--755",
isbn="978-3-319-10602-1"
}

@ARTICLE{ramulator2015,
  author={Kim, Yoongu and Yang, Weikun and Mutlu, Onur},
  journal={IEEE Computer Architecture Letters}, 
  title={{Ramulator: A Fast and Extensible DRAM Simulator}}, 
  year={2016},
  volume={15},
  number={1},
  pages={45-49}}

@inproceedings{Ahn2015isca,
author = {Ahn, Junwhan and Yoo, Sungjoo and Mutlu, Onur and Choi, Kiyoung},
title = {{PIM-enabled instructions: a low-overhead, locality-aware processing-in-memory architecture}},
year = {2015},
isbn = {9781450334020},
publisher = {Association for Computing Machinery},
address = {New York, NY, USA},
url = {https://doi.org/10.1145/2749469.2750385},
doi = {10.1145/2749469.2750385},
booktitle = {Proceedings of the 42nd Annual International Symposium on Computer Architecture},
pages = {336–348},
numpages = {13},
location = {Portland, Oregon},
series = {ISCA '15}
}

@article{kitti2013vision,
  title={Vision meets robotics: The kitti dataset},
  author={Geiger, Andreas and Lenz, Philip and Stiller, Christoph and Urtasun, Raquel},
  journal={The international journal of robotics research},
  volume={32},
  number={11},
  pages={1231--1237},
  year={2013},
  publisher={Sage Publications Sage UK: London, England}
}

@InProceedings{Dai2017ICCV,
author = {Dai, Jifeng and Qi, Haozhi and Xiong, Yuwen and Li, Yi and Zhang, Guodong and Hu, Han and Wei, Yichen},
title = {{Deformable Convolutional Networks}},
booktitle = {Proceedings of the IEEE International Conference on Computer Vision (ICCV)},
month = {Oct},
year = {2017}}

@inproceedings{Vaswani17,
  title={{Attention is all you need}},
  author={Vaswani, Ashish and Shazeer, Noam and Parmar, Niki and Uszkoreit, Jakob and Jones, Llion and Gomez, Aidan N and Kaiser, {\L}ukasz and Polosukhin, Illia},
  booktitle={Advances in Neural Information Processing Systems},
  pages={5998--6008},
  year={2017}
}

@article{bert18,
  author    = {Jacob Devlin and
               Ming{-}Wei Chang and
               Kenton Lee and
               Kristina Toutanova},
  title     = {{BERT:} Pre-training of Deep Bidirectional Transformers for Language
               Understanding},
  journal   = {CoRR},
  volume    = {abs/1810.04805},
  year      = {2018},
  archivePrefix = {arXiv},
  eprint    = {1810.04805}
}

@inproceedings{Aken19,
author = {van Aken, Betty and Winter, Benjamin and L\"{o}ser, Alexander and Gers, Felix A.},
title = {{How Does BERT Answer Questions? A Layer-Wise Analysis of Transformer Representations}},
year = {2019},
isbn = {9781450369763},
publisher = {Association for Computing Machinery},
address = {New York, NY, USA},
url = {https://doi.org/10.1145/3357384.3358028},
doi = {10.1145/3357384.3358028},
booktitle = {Proceedings of the 28th ACM International Conference on Information and Knowledge Management},
pages = {1823–1832},
numpages = {10},
location = {Beijing, China},
series = {CIKM '19}
}

@article{RoBERTa2019,
  author    = {Yinhan Liu and
               Myle Ott and
               Naman Goyal and
               Jingfei Du and
               Mandar Joshi and
               Danqi Chen and
               Omer Levy and
               Mike Lewis and
               Luke Zettlemoyer and
               Veselin Stoyanov},
  title     = {{RoBERTa: A Robustly Optimized BERT Pretraining Approach}},
  journal   = {CoRR},
  volume    = {abs/1907.11692},
  year      = {2019},
  url       = {http://arxiv.org/abs/1907.11692},
  archivePrefix = {arXiv},
  eprint    = {1907.11692},
  bibsource = {dblp computer science bibliography, https://dblp.org}
}

@article{longformer20,
  author    = {Iz Beltagy and
               Matthew E. Peters and
               Arman Cohan},
  title     = {Longformer: The Long-Document Transformer},
  journal   = {CoRR},
  volume    = {abs/2004.05150},
  year      = {2020},
  url       = {https://arxiv.org/abs/2004.05150},
  eprinttype = {arXiv},
  eprint    = {2004.05150},
  bibsource = {dblp computer science bibliography, https://dblp.org}
}

@INPROCEEDINGS{recnmp2020,
  author={Ke, Liu and Gupta, Udit and Cho, Benjamin Youngjae and Brooks, David and Chandra, Vikas and Diril, Utku and Firoozshahian, Amin and Hazelwood, Kim and Jia, Bill and Lee, Hsien-Hsin S. and Li, Meng and Maher, Bert and Mudigere, Dheevatsa and Naumov, Maxim and Schatz, Martin and Smelyanskiy, Mikhail and Wang, Xiaodong and Reagen, Brandon and Wu, Carole-Jean and Hempstead, Mark and Zhang, Xuan},
  booktitle={2020 ACM/IEEE 47th Annual International Symposium on Computer Architecture (ISCA)}, 
  title={{RecNMP: Accelerating Personalized Recommendation with Near-Memory Processing}}, 
  year={2020},
  volume={},
  number={},
  pages={790-803}}

@inproceedings{Tay20,
  title={Sparse sinkhorn attention},
  author={Tay, Yi and Bahri, Dara and Yang, Liu and Metzler, Donald and Juan, Da-Cheng},
  booktitle={International Conference on Machine Learning (ICML)},
  pages={9438--9447},
  year={2020},
  organization={PMLR}
}

@inproceedings{Zaheer20,
  title={{Big Bird}: Transformers for Longer Sequences},
  author={Zaheer, Manzil and Guruganesh, Guru and Dubey, Kumar Avinava and Ainslie, Joshua and Alberti, Chris and Ontanon, Santiago and Pham, Philip and Ravula, Anirudh and Wang, Qifan and Yang, Li and Ahmed, Amr},
  booktitle={Advances in Neural Information Processing Systems (NeurIPS)},
  volume={33},
  pages={17283--17297},
  year={2020}
}

@inproceedings{Li20,
author = {Li, Bingbing and Pandey, Santosh and Fang, Haowen and Lyv, Yanjun and Li, Ji and Chen, Jieyang and Xie, Mimi and Wan, Lipeng and Liu, Hang and Ding, Caiwen},
title = {{FTRANS}: Energy-Efficient Acceleration of Transformers Using {FPGA}},
year = {2020},
isbn = {9781450370530},
publisher = {Association for Computing Machinery},
address = {New York, NY, USA},
url = {https://doi.org/10.1145/3370748.3406567},
doi = {10.1145/3370748.3406567},
booktitle = {Proceedings of the ACM/IEEE International Symposium on Low Power Electronics and Design},
pages = {175–180},
numpages = {6},
location = {Boston, Massachusetts},
series = {ISLPED '20}
}

@article{Performer2020,
  author       = {Krzysztof Choromanski and
                  Valerii Likhosherstov and
                  David Dohan and
                  Xingyou Song and
                  Andreea Gane and
                  Tam{\'{a}}s Sarl{\'{o}}s and
                  Peter Hawkins and
                  Jared Davis and
                  Afroz Mohiuddin and
                  Lukasz Kaiser and
                  David Belanger and
                  Lucy J. Colwell and
                  Adrian Weller},
  title        = {Rethinking Attention with Performers},
  journal      = {CoRR},
  volume       = {abs/2009.14794},
  year         = {2020},
  url          = {https://arxiv.org/abs/2009.14794},
  eprinttype    = {arXiv},
  eprint       = {2009.14794},
  bibsource    = {dblp computer science bibliography, https://dblp.org}
}

@article{Linformer2020,
  author       = {Sinong Wang and
                  Belinda Z. Li and
                  Madian Khabsa and
                  Han Fang and
                  Hao Ma},
  title        = {Linformer: Self-Attention with Linear Complexity},
  journal      = {CoRR},
  volume       = {abs/2006.04768},
  year         = {2020},
  url          = {https://arxiv.org/abs/2006.04768},
  eprinttype    = {arXiv},
  eprint       = {2006.04768},
  bibsource    = {dblp computer science bibliography, https://dblp.org}
}

@article{wu2020visual,
      title={{Visual Transformers: Token-based Image Representation and Processing for Computer Vision}}, 
      author={Bichen Wu and Chenfeng Xu and Xiaoliang Dai and Alvin Wan and Peizhao Zhang and Zhicheng Yan and Masayoshi Tomizuka and Joseph Gonzalez and Kurt Keutzer and Peter Vajda},
      year={2020},
      journal={arXiv preprint arXiv:2006.03677},
      eprint={2006.03677},
      archivePrefix={arXiv},
      primaryClass={cs.CV}
}

@inproceedings{Yang20,
  title={{ReTransformer}: {ReRAM}-based processing-in-memory architecture for transformer acceleration},
  author={Yang, Xiaoxuan and Yan, Bonan and Li, Hai and Chen, Yiran},
  booktitle={Proceedings of the 39th International Conference on Computer-Aided Design (ICCAD)},
  pages={1--9},
  year={2020}
}

@InProceedings{Arnab_2021_ICCV,
    author    = {Arnab, Anurag and Dehghani, Mostafa and Heigold, Georg and Sun, Chen and Lu\v{c}i\'c, Mario and Schmid, Cordelia},
    title     = {{ViViT: A Video Vision Transformer}},
    booktitle = {Proceedings of the IEEE/CVF International Conference on Computer Vision (ICCV)},
    month     = {October},
    year      = {2021},
    pages     = {6836-6846}
}

@article{Fang2021,
  author    = {Yuxin Fang and
               Bencheng Liao and
               Xinggang Wang and
               Jiemin Fang and
               Jiyang Qi and
               Rui Wu and
               Jianwei Niu and
               Wenyu Liu},
  title     = {{You Only Look at One Sequence: Rethinking Transformer in Vision through
               Object Detectio}},
  journal   = {CoRR},
  volume    = {abs/2106.00666},
  year      = {2021},
  url       = {https://arxiv.org/abs/2106.00666},
  eprinttype = {arXiv},
  eprint    = {2106.00666},
  bibsource = {dblp computer science bibliography, https://dblp.org}
}

@INPROCEEDINGS{Kal2021,
  author={Kal, Hongju and Lee, Seokmin and Ko, Gun and Ro, Won Woo},
  booktitle={2021 ACM/IEEE 48th Annual International Symposium on Computer Architecture (ISCA)}, 
  title={{SPACE: Locality-Aware Processing in Heterogeneous Memory for Personalized Recommendations}}, 
  year={2021},
  volume={},
  number={},
  pages={679-691},
  doi={10.1109/ISCA52012.2021.00059}}

@ARTICLE{Kang21,
  author={Kang, Myeonggu and Shin, Hyein and Kim, Lee-Sup},
  journal={IEEE Transactions on Computer-Aided Design of Integrated Circuits and Systems}, 
  title={{A Framework for Accelerating Transformer-based Language Model on ReRAM-based Architecture}}, 
  year={2021},
  volume={},
  number={},
  pages={1-1},
  doi={10.1109/TCAD.2021.3121264}}

@INPROCEEDINGS{Laguna2021,
  author={Laguna, Ann Franchesca and Kazemi, Arman and Niemier, Michael and Hu, X. Sharon},
  booktitle={2021 Design, Automation \& Test in Europe Conference \& Exhibition (DATE)}, 
  title={{In-Memory Computing based Accelerator for Transformer Networks for Long Sequences}}, 
  year={2021},
  volume={},
  number={},
  pages={1839-1844},
  doi={10.23919/DATE51398.2021.9474146}}

@INPROCEEDINGS{Zhou2021,
  author={Zhou, Minxuan and Guo, Yunhui and Xu, Weihong and Li, Bin and Eliceiri, Kevin W. and Rosing, Tajana},
  booktitle={2021 58th ACM/IEEE Design Automation Conference (DAC)}, 
  title={{MAT: Processing In-Memory Acceleration for Long-Sequence Attention}}, 
  year={2021},
  volume={},
  number={},
  pages={25-30},
  doi={10.1109/DAC18074.2021.9586212}}

@ARTICLE{Kim2022,
  author={Kim, Jin Hyun and Kang, Shin-Haeng and Lee, Sukhan and Kim, Hyeonsu and Ro, Yuhwan and Lee, Seungwon and Wang, David and Choi, Jihyun and So, Jinin and Cho, YeonGon and Song, JoonHo and Cho, Jeonghyeon and Sohn, Kyomin and Kim, Nam Sung},
  journal={IEEE Micro}, 
  title={{Aquabolt-XL HBM2-PIM, LPDDR5-PIM With In-Memory Processing, and AXDIMM With Acceleration Buffer}}, 
  year={2022},
  volume={42},
  number={3},
  pages={20-30},
  doi={10.1109/MM.2022.3164651}}

@INPROCEEDINGS{sprint2022,
  author={Yazdanbakhsh, Amir and Moradifirouzabadi, Ashkan and Li, Zheng and Kang, Mingu},
  booktitle={Proceedings of 2022 55th IEEE/ACM International Symposium on Microarchitecture (MICRO)}, 
  title={{Sparse Attention Acceleration with Synergistic In-Memory Pruning and On-Chip Recomputation}}, 
  year={2022},
  volume={},
  number={},
  pages={744-762},
  doi={10.1109/MICRO56248.2022.00059}}

@inproceedings{Liu2023,
author = {Liu, Haifeng and Zheng, Long and Huang, Yu and Liu, Chaoqiang and Ye, Xiangyu and Yuan, Jingrui and Liao, Xiaofei and Jin, Hai and Xue, Jingling},
title = {{Accelerating Personalized Recommendation with Cross-Level Near-Memory Processing}},
year = {2023},
isbn = {9798400700958},
publisher = {Association for Computing Machinery},
address = {New York, NY, USA},
url = {https://doi.org/10.1145/3579371.3589101},
doi = {10.1145/3579371.3589101},
booktitle = {Proceedings of the 50th Annual International Symposium on Computer Architecture},
articleno = {66},
numpages = {13},
location = {Orlando, FL, USA},
series = {ISCA '23}
}

@ARTICLE{Li2023,
  author={Li, Wantong and Manley, Madison and Read, James and Kaul, Ankit and Bakir, Muhannad S. and Yu, Shimeng},
  journal={IEEE Transactions on Very Large Scale Integration (VLSI) Systems}, 
  title={{H3DAtten: Heterogeneous 3-D Integrated Hybrid Analog and Digital Compute-in-Memory Accelerator for Vision Transformer Self-Attention}}, 
  year={2023},
  volume={31},
  number={10},
  pages={1592-1602},
  doi={10.1109/TVLSI.2023.3299509}}

@inproceedings{Qin2023,
author = {Qin, Yubin and Wang, Yang and Deng, Dazheng and Zhao, Zhiren and Yang, Xiaolong and Liu, Leibo and Wei, Shaojun and Hu, Yang and Yin, Shouyi},
title = {{FACT: FFN-Attention Co-Optimized Transformer Architecture with Eager Correlation Prediction}},
year = {2023},
isbn = {9798400700958},
publisher = {Association for Computing Machinery},
address = {New York, NY, USA},
url = {https://doi.org/10.1145/3579371.3589057},
doi = {10.1145/3579371.3589057},
booktitle = {Proceedings of the 50th Annual International Symposium on Computer Architecture},
articleno = {22},
numpages = {14},
location = {Orlando, FL, USA},
series = {ISCA '23}
}

@ARTICLE{Sridharan2023,
  author={Sridharan, Shrihari and Stevens, Jacob R. and Roy, Kaushik and Raghunathan, Anand},
  journal={IEEE Transactions on Very Large Scale Integration (VLSI) Systems}, 
  title={{X-Former: In-Memory Acceleration of Transformers}}, 
  year={2023},
  volume={31},
  number={8},
  pages={1223-1233},
  doi={10.1109/TVLSI.2023.3282046}}

@article{Ye2023,
author = {Ye, Wenhua and Zhou, Xu and Zhou, Joey and Chen, Cen and Li, Kenli},
title = {Accelerating Attention Mechanism on FPGAs Based on Efficient Reconfigurable Systolic Array},
year = {2023},
issue_date = {November 2023},
publisher = {Association for Computing Machinery},
address = {New York, NY, USA},
volume = {22},
number = {6},
issn = {1539-9087},
url = {https://doi.org/10.1145/3549937},
doi = {10.1145/3549937},
journal = {ACM Trans. Embed. Comput. Syst.},
month = {nov},
articleno = {93},
numpages = {22}
}

@INPROCEEDINGS{You2023,
  author={You, Haoran and Sun, Zhanyi and Shi, Huihong and Yu, Zhongzhi and Zhao, Yang and Zhang, Yongan and Li, Chaojian and Li, Baopu and Lin, Yingyan},
  booktitle={2023 IEEE International Symposium on High-Performance Computer Architecture (HPCA)}, 
  title={{ViTCoD: Vision Transformer Acceleration via Dedicated Algorithm and Accelerator Co-Design}}, 
  year={2023},
  volume={},
  number={},
  pages={273-286},
  doi={10.1109/HPCA56546.2023.10071027}}

@INPROCEEDINGS{Zheng2023,
  author={Zheng, Qilin and Li, Shiyu and Wang, Yitu and Li, Ziru and Chen, Yiran and Li, Hai Helen},
  booktitle={2023 60th ACM/IEEE Design Automation Conference (DAC)}, 
  title={{Accelerating Sparse Attention with a Reconfigurable Non-volatile Processing-In-Memory Architecture}}, 
  year={2023},
  volume={},
  number={},
  pages={1-6},
  doi={10.1109/DAC56929.2023.10247908}}

@inproceedings{Angizi2018,  
 title={{CMP-PIM: an energy-efficient comparator-based processing-in-memory neural network accelerator}}, 
 url={http://dx.doi.org/10.1145/3195970.3196009}, 
 DOI={10.1145/3195970.3196009}, 
 booktitle={Proceedings of the 55th Annual Design Automation Conference}, 
 author={Angizi, Shaahin and He, Zhezhi and Rakin, Adnan Siraj and Fan, Deliang}, 
 year={2018}, 
 month={Jun}, 
 language={en-US} 
 }

@inproceedings{Lu21,
author = {Lu, Liqiang and Jin, Yicheng and Bi, Hangrui and Luo, Zizhang and Li, Peng and Wang, Tao and Liang, Yun},
title = {Sanger: A Co-Design Framework for Enabling Sparse Attention Using Reconfigurable Architecture},
year = {2021},
isbn = {9781450385572},
publisher = {Association for Computing Machinery},
address = {New York, NY, USA},
doi = {10.1145/3466752.3480125},
booktitle = {Proceedings of 54th Annual IEEE/ACM International Symposium on Microarchitecture},
pages = {977–991},
numpages = {15},
location = {Virtual Event, Greece},
}

@inproceedings{Park2021,
author = {Park, Jaehyun and Kim, Byeongho and Yun, Sungmin and Lee, Eojin and Rhu, Minsoo and Ahn, Jung Ho},
title = {{TRiM: Enhancing Processor-Memory Interfaces with Scalable Tensor Reduction in Memory}},
year = {2021},
booktitle = {MICRO-54: 54th Annual IEEE/ACM International Symposium on Microarchitecture},
pages = {268–281},
numpages = {14}
}

@inproceedings{zhu2021deformable,
title={{Deformable DETR: Deformable Transformers for End-to-End Object Detection}},
author={Xizhou Zhu and Weijie Su and Lewei Lu and Bin Li and Xiaogang Wang and Jifeng Dai},
booktitle={International Conference on Learning Representations},
year={2021},
url={https://openreview.net/forum?id=gZ9hCDWe6ke}
}

@article{Zhang21fpga,
author = {Xinyi Zhang and Yawen Wu and Peipei Zhou and Xulong Tang and Jingtong Hu},
title = {Algorithm-Hardware Co-Design of Attention Mechanism on {FPGA} Devices},
journal = {ACM Trans. Embed. Comput. Syst.},
volume={20},
number={5s},
pages={1--24},
year={2021}
}

@article{chen2016eyeriss,
  title={Eyeriss: A spatial architecture for energy-efficient dataflow for convolutional neural networks},
  author={Chen, Yu-Hsin and Emer, Joel and Sze, Vivienne},
  journal={ACM SIGARCH computer architecture news},
  volume={44},
  number={3},
  pages={367--379},
  year={2016},
  publisher={ACM New York, NY, USA}
}

@inproceedings{kao2023flat,
  title={Flat: An optimized dataflow for mitigating attention bottlenecks},
  author={Kao, Sheng-Chun and Subramanian, Suvinay and Agrawal, Gaurav and Yazdanbakhsh, Amir and Krishna, Tushar},
  booktitle={Proceedings of the 28th ACM International Conference on Architectural Support for Programming Languages and Operating Systems, Volume 2},
  pages={295--310},
  year={2023}
}

@inproceedings{lee2024infinigen,
  title={$\{$InfiniGen$\}$: Efficient generative inference of large language models with dynamic $\{$KV$\}$ cache management},
  author={Lee, Wonbeom and Lee, Jungi and Seo, Junghwan and Sim, Jaewoong},
  booktitle={18th USENIX Symposium on Operating Systems Design and Implementation (OSDI 24)},
  pages={155--172},
  year={2024}
}

@article{gilbert2024looptree,
  title={LoopTree: Exploring the Fused-layer Dataflow Accelerator Design Space},
  author={Gilbert, Michael and Wu, Yannan Nellie and Emer, Joel S and Sze, Vivienne},
  journal={IEEE Transactions on Circuits and Systems for Artificial Intelligence},
  year={2024},
  publisher={IEEE}
}

@INPROCEEDINGS{Fan2022fpga,
  author={Fan, Hongxiang and Chau, Thomas and Venieris, Stylianos I. and Lee, Royson and Kouris, Alexandros and Luk, Wayne and Lane, Nicholas D. and Abdelfattah, Mohamed S.},
  booktitle={2022 55th IEEE/ACM International Symposium on Microarchitecture (MICRO)}, 
  title={{Adaptable Butterfly Accelerator for Attention-based NNs via Hardware and Algorithm Co-design}}, 
  year={2022},
  volume={},
  number={},
  pages={599-615},
  doi={10.1109/MICRO56248.2022.00050}}

@inproceedings{Flashatten2022,
 author = {Dao, Tri and Fu, Dan and Ermon, Stefano and Rudra, Atri and R\'{e}, Christopher},
 booktitle = {Advances in Neural Information Processing Systems},
 pages = {16344--16359},
 title = {{FlashAttention: Fast and Memory-Efficient Exact Attention with IO-Awareness}},
 volume = {35},
 year = {2022}
}

@inproceedings{wang2020figaro,
  title={FIGARO: Improving system performance via fine-grained in-DRAM data relocation and caching},
  author={Wang, Yaohua and Orosa, Lois and Peng, Xiangjun and Guo, Yang and Ghose, Saugata and Patel, Minesh and Kim, Jeremie S and Luna, Juan G{\'o}mez and Sadrosadati, Mohammad and Ghiasi, Nika Mansouri and others},
  booktitle={2020 53rd Annual IEEE/ACM International Symposium on Microarchitecture (MICRO)},
  pages={313--328},
  year={2020},
  organization={IEEE}
}

@inproceedings{chang2016low,
  title={Low-cost inter-linked subarrays (LISA): Enabling fast inter-subarray data movement in DRAM},
  author={Chang, Kevin K and Nair, Prashant J and Lee, Donghyuk and Ghose, Saugata and Qureshi, Moinuddin K and Mutlu, Onur},
  booktitle={2016 IEEE International Symposium on High Performance Computer Architecture (HPCA)},
  pages={568--580},
  year={2016},
  organization={IEEE}
}

@inproceedings{hassan2019crow,
  title={Crow: A low-cost substrate for improving dram performance, energy efficiency, and reliability},
  author={Hassan, Hasan and Patel, Minesh and Kim, Jeremie S and Yaglikci, A Giray and Vijaykumar, Nandita and Ghiasi, Nika Mansouri and Ghose, Saugata and Mutlu, Onur},
  booktitle={Proceedings of the 46th International Symposium on Computer Architecture},
  pages={129--142},
  year={2019}
}

@InProceedings{Li_2022_CVPR,
    author    = {Li, Feng and Zhang, Hao and Liu, Shilong and Guo, Jian and Ni, Lionel M. and Zhang, Lei},
    title     = {{DN-DETR: Accelerate DETR Training by Introducing Query DeNoising}},
    booktitle = {Proceedings of the IEEE/CVF Conference on Computer Vision and Pattern Recognition (CVPR)},
    month     = {June},
    year      = {2022},
    pages     = {13619-13627}
}

@inproceedings{Qu22,
author = {Qu, Zheng and Liu, Liu and Tu, Fengbin and Chen, Zhaodong and Ding, Yufei and Xie, Yuan},
title = {DOTA: Detect and Omit Weak Attentions for Scalable Transformer Acceleration},
year = {2022},
isbn = {9781450392051},
publisher = {Association for Computing Machinery},
doi = {10.1145/3503222.3507738},
booktitle = {Proceedings of the 27th ACM International Conference on Architectural Support for Programming Languages and Operating Systems},
pages = {14–26},
numpages = {13},
location = {Lausanne, Switzerland},
}

@inproceedings{
zhang2023dino,
title={{DINO}: {DETR} with Improved DeNoising Anchor Boxes for End-to-End Object Detection},
author={Hao Zhang and Feng Li and Shilong Liu and Lei Zhang and Hang Su and Jun Zhu and Lionel Ni and Heung-Yeung Shum},
booktitle={The Eleventh International Conference on Learning Representations },
year={2023},
url={https://openreview.net/forum?id=3mRwyG5one}
}

@INPROCEEDINGS{Zhou2022,
  author={Zhou, Minxuan and Xu, Weihong and Kang, Jaeyoung and Rosing, Tajana},
  booktitle={Proceedings of 2022 IEEE International Symposium on High-Performance Computer Architecture (HPCA)}, 
  title={{TransPIM: A Memory-based Acceleration via Software-Hardware Co-Design for Transformer}}, 
  year={2022},
  volume={},
  number={},
  pages={1071-1085}}

@inproceedings{chen2023,
author = {Chen, Dan and He, Haiheng and Jin, Hai and Zheng, Long and Huang, Yu and Shen, Xinyang and Liao, Xiaofei},
title = {MetaNMP: Leveraging Cartesian-Like Product to Accelerate HGNNs with Near-Memory Processing},
year = {2023},
isbn = {9798400700958},
publisher = {Association for Computing Machinery},
address = {New York, NY, USA},
url = {https://doi.org/10.1145/3579371.3589091},
doi = {10.1145/3579371.3589091},
booktitle = {Proceedings of the 50th Annual International Symposium on Computer Architecture},
articleno = {56},
numpages = {13},
location = {Orlando, FL, USA},
series = {ISCA '23}
}

@INPROCEEDINGS{Ding2023,
  author={Ding, Yan and Liu, Chubo and Duan, Mingxing and Chang, Wanli and Li, Keqin and Li, Kenli},
  booktitle={2023 60th ACM/IEEE Design Automation Conference (DAC)}, 
  title={{HAIMA: A Hybrid SRAM and DRAM Accelerator-in-Memory Architecture for Transformer}}, 
  year={2023},
  volume={},
  number={},
  pages={1-6}}

@inproceedings{Qin23isca,
author = {Qin, Yubin and Wang, Yang and Deng, Dazheng and Zhao, Zhiren and Yang, Xiaolong and Liu, Leibo and Wei, Shaojun and Hu, Yang and Yin, Shouyi},
title = {{FACT: FFN-Attention Co-optimized Transformer Architecture with Eager Correlation Prediction}},
year = {2023},
isbn = {9798400700958},
publisher = {Association for Computing Machinery},
address = {New York, NY, USA},
url = {https://doi.org/10.1145/3579371.3589057},
doi = {10.1145/3579371.3589057},
articleno = {22},
numpages = {14},
location = {Orlando, FL, USA},
series = {ISCA '23}
}

@article{Zhang21,
author = {Xinyi Zhang and Yawen Wu and Peipei Zhou and Xulong Tang and Jingtong Hu},
title = {{Algorithm-Hardware Co-Design of Attention Mechanism on FPGA Devices}},
journal = {ACM Trans. Embed. Comput. Syst.},
volume={20},
number={5s},
pages={1--24},
year={2021}
}

@INPROCEEDINGS{Ham20,
  author={Ham, Tae Jun and Jung, Sung Jun and Kim, Seonghak and Oh, Young H. and Park, Yeonhong and Song, Yoonho and Park, Jung-Hun and Lee, Sanghee and Park, Kyoung and Lee, Jae W. and Jeong, Deog-Kyoon},
  booktitle={Proceedings of 2020 IEEE International Symposium on High Performance Computer Architecture (HPCA)}, 
  title={A$^3$: Accelerating Attention Mechanisms in Neural Networks with Approximation}, 
  year={2020},
  volume={},
  number={},
  pages={328-341}}

@inproceedings{Gao2017,  
 title={{TETRIS: Scalable and Efficient Neural Network Acceleration with 3D Memory}}, 
 booktitle={Proceedings of the Twenty-Second International Conference on Architectural Support for Programming Languages and Operating Systems}, 
 author={Gao, Mingyu and Pu, Jing and Yang, Xuan and Horowitz, Mark and Kozyrakis, Christos}, 
 year={2017}, 
 month={Apr}, 
 language={en-US} 
 }

@inproceedings{Zhang2018,  
 title={{GraphP: Reducing Communication for PIM-Based Graph Processing with Efficient Data Partition}}, 
 booktitle={2018 IEEE International Symposium on High Performance Computer Architecture (HPCA)}, 
 author={Zhang, Mingxing and Zhuo, Youwei and Wang, Chao and Gao, Mingyu and Wu, Yongwei and Chen, Kang and Kozyrakis, Christos and Qian, Xuehai}, 
 year={2018}, 
 month={Feb}, 
 language={en-US} 
 }

@inproceedings{Dai2022,
author = {Dai, Guohao and Zhu, Zhenhua and Fu, Tianyu and Wei, Chiyue and Wang, Bangyan and Li, Xiangyu and Xie, Yuan and Yang, Huazhong and Wang, Yu},
title = {{DIMMining: Pruning-Efficient and Parallel Graph Mining on near-Memory-Computing}},
year = {2022},
booktitle = {Proceedings of the 49th Annual International Symposium on Computer Architecture},
pages = {130–145},
numpages = {16}
}

@inproceedings{Talati2022,
author = {Talati, Nishil and Ye, Haojie and Yang, Yichen and Belayneh, Leul and Chen, Kuan-Yu and Blaauw, David and Mudge, Trevor and Dreslinski, Ronald},
title = {{NDMiner: Accelerating Graph Pattern Mining Using near Data Processing}},
year = {2022},
booktitle = {Proceedings of the 49th Annual International Symposium on Computer Architecture},
pages = {146–159},
numpages = {14}
}

@article{Xu2024,
author = {Xu, Jiahong and Liu, Haikun and Duan, Zhuohui and Liao, Xiaofei and Jin, Hai and Yang, Xiaokang and Li, Huize and Liu, Cong and Mao, Fubing and Zhang, Yu},
title = {{ReHarvest: An ADC Resource-Harvesting Crossbar Architecture for ReRAM-Based DNN Accelerators}},
year = {2024},
publisher = {Association for Computing Machinery},
address = {New York, NY, USA},
volume = {21},
number = {3},
journal = {ACM Trans. Archit. Code Optim.},
month = sep,
articleno = {63},
numpages = {26}
}

@inproceedings{ReGNN2022,
author = {Liu, Cong and Liu, Haikun and Jin, Hai and Liao, Xiaofei and Zhang, Yu and Duan, Zhuohui and Xu, Jiahong and Li, Huize},
title = {{ReGNN: a ReRAM-based heterogeneous architecture for general graph neural networks}},
year = {2022},
booktitle = {Proceedings of the 59th ACM/IEEE Design Automation Conference},
pages = {469–474},
numpages = {6}
}

@INPROCEEDINGS{Xie2021,
  author={Xie, Xinfeng and Liang, Zheng and Gu, Peng and Basak, Abanti and Deng, Lei and Liang, Ling and Hu, Xing and Xie, Yuan},
  booktitle={2021 IEEE International Symposium on High-Performance Computer Architecture (HPCA)}, 
  title={{SpaceA: Sparse Matrix Vector Multiplication on Processing-in-Memory Accelerator}}, 
  year={2021},
  volume={},
  number={},
  pages={570-583}}

@ARTICLE{li2022cpsaa,
  author={Li, Huize and Jin, Hai and Zheng, Long and Liao, Xiaofei and Huang, Yu and Liu, Cong and Xu, Jiahong and Duan, Zhuohui and Chen, Dan and Gui, Chuangyi},
  journal={IEEE Transactions on Computer-Aided Design of Integrated Circuits and Systems}, 
  title={{CPSAA: Accelerating Sparse Attention Using Crossbar-Based Processing-In-Memory Architecture}}, 
  year={2023},
  volume={},
  number={},
  pages={1-1}}

@INPROCEEDINGS{asadi2024,
  author={Li, Huize and Li, Zhaoying and Bai, Zhenyu and Mitra, Tulika},
  booktitle={2024 IEEE International Symposium on High-Performance Computer Architecture (HPCA)}, 
  title={{ASADI: Accelerating Sparse Attention Using Diagonal-based In-Situ Computing}}, 
  year={2024},
  volume={},
  number={},
  pages={774-787}}

@inproceedings{bai2024dac,
author = {Bai, Zhenyu and Dangi, Pranav and Li, Huize and Mitra, Tulika},
title = {{SWAT: Scalable and Efficient Window Attention-based Transformers Acceleration on FPGAs}},
year = {2024},
isbn = {9798400706011},
publisher = {Association for Computing Machinery},
address = {New York, NY, USA},
url = {https://doi.org/10.1145/3649329.3658488},
doi = {10.1145/3649329.3658488},
booktitle = {Proceedings of the 61st ACM/IEEE Design Automation Conference},
articleno = {93},
numpages = {6},
location = {San Francisco, CA, USA},
series = {DAC '24}
}

@inproceedings{xu2024defa,
author = {Xu, Yansong and Lyu, Dongxu and Li, Zhenyu and Chen, Yuzhou and Wang, Zilong and Wang, Gang and Wang, Zhican and Li, Haomin and He, Guanghui},
title = {{DEFA: Efficient Deformable Attention Acceleration via Pruning-Assisted Grid-Sampling and Multi-Scale Parallel Processing}},
year = {2024},
isbn = {9798400706011},
publisher = {Association for Computing Machinery},
address = {New York, NY, USA},
url = {https://doi.org/10.1145/3649329.3657328},
doi = {10.1145/3649329.3657328},
booktitle = {Proceedings of the 61st ACM/IEEE Design Automation Conference},
articleno = {24},
numpages = {6},
location = {San Francisco, CA, USA},
series = {DAC '24}
}

@ARTICLE{sadimm2025,
  author={Li, Huize and Chen, Dan and Mitra, Tulika},
  journal={IEEE Transactions on Computers}, 
  title={{SADIMM: Accelerating Sparse Attention Using DIMM-Based Near-Memory Processing}}, 
  year={2025},
  volume={74},
  number={2},
  pages={542-554},
  doi={10.1109/TC.2024.3500362}}

@inproceedings{carion2020end,
  title={End-to-end object detection with transformers},
  author={Carion, Nicolas and Massa, Francisco and Synnaeve, Gabriel and Usunier, Nicolas and Kirillov, Alexander and Zagoruyko, Sergey},
  booktitle={European conference on computer vision},
  pages={213--229},
  year={2020},
  organization={Springer}
}

@inproceedings{gupta2022ow,
  title={Ow-detr: Open-world detection transformer},
  author={Gupta, Akshita and Narayan, Sanath and Joseph, KJ and Khan, Salman and Khan, Fahad Shahbaz and Shah, Mubarak},
  booktitle={Proceedings of the IEEE/CVF conference on computer vision and pattern recognition},
  pages={9235--9244},
  year={2022}
}

@article{beal2020toward,
  title={Toward transformer-based object detection},
  author={Beal, Josh and Kim, Eric and Tzeng, Eric and Park, Dong Huk and Zhai, Andrew and Kislyuk, Dmitry},
  journal={arXiv preprint arXiv:2012.09958},
  year={2020}
}

@inproceedings{du2018understanding,
  title={Understanding of object detection based on CNN family and YOLO},
  author={Du, Juan},
  booktitle={Journal of Physics: Conference Series},
  volume={1004},
  pages={012029},
  year={2018},
  organization={IOP Publishing}
}

@article{Rajpurkar18,
  title={{Know what you don't know: Unanswerable questions for SQuAD}},
  author={Rajpurkar, Pranav and Jia, Robin and Liang, Percy},
  journal={arXiv preprint arXiv:1806.03822},
  year={2018}
}

@inproceedings{xie2021oriented,
  title={Oriented R-CNN for object detection},
  author={Xie, Xingxing and Cheng, Gong and Wang, Jiabao and Yao, Xiwen and Han, Junwei},
  booktitle={Proceedings of the IEEE/CVF international conference on computer vision},
  pages={3520--3529},
  year={2021}
}

@inproceedings{chen2016r,
  title={R-CNN for small object detection},
  author={Chen, Chenyi and Liu, Ming-Yu and Tuzel, Oncel and Xiao, Jianxiong},
  booktitle={Asian conference on computer vision},
  pages={214--230},
  year={2016},
  organization={Springer}
}

\end{document}